\documentstyle[12pt,epsf]{article}

  \def\pp{{\mathchoice
          {%displaystyle
              \kern 1pt%
              \raise 1pt
              \vbox{\hrule width5pt height0.4pt depth0pt
                    \kern -2pt
                    \hbox{\kern 2.3pt
                          \vrule width0.4pt height6pt depth0pt
                          }
                    \kern -2pt
                    \hrule width5pt height0.4pt depth0pt}%
                    \kern 1pt
           }
            {%textstyle
              \kern 1pt%
              \raise 1pt
              \vbox{\hrule width4.3pt height0.4pt depth0pt
                    \kern -1.8pt
                    \hbox{\kern 1.95pt
                          \vrule width0.4pt height5.4pt depth0pt
                          }
                    \kern -1.8pt
                    \hrule width4.3pt height0.4pt depth0pt}%
                    \kern 1pt
            }
            {%scriptstyle
              \kern 0.5pt%
              \raise 1pt
              \vbox{\hrule width4.0pt height0.3pt depth0pt
                    \kern -1.9pt  %[e]=0.15pt
                    \hbox{\kern 1.85pt
                          \vrule width0.3pt height5.7pt depth0pt
                          }
                    \kern -1.9pt
                    \hrule width4.0pt height0.3pt depth0pt}%
                    \kern 0.5pt
            }
            {%scriptscriptstyle
              \kern 0.5pt%
              \raise 1pt
              \vbox{\hrule width3.6pt height0.3pt depth0pt
                    \kern -1.5pt
                    \hbox{\kern 1.65pt
                          \vrule width0.3pt height4.5pt depth0pt
                          }
                    \kern -1.5pt
                    \hrule width3.6pt height0.3pt depth0pt}%
                    \kern 0.5pt%}
            }
        }}

  \def\mm{{\mathchoice
                       {%displaystyle
                             \kern 1pt
               \raise 1pt    \vbox{\hrule width5pt height0.4pt depth0pt
                                  \kern 2pt
                                  \hrule width5pt height0.4pt depth0pt}
                             \kern 1pt}
                       {%textstyle
                            \kern 1pt
               \raise 1pt \vbox{\hrule width4.3pt height0.4pt depth0pt
                                  \kern 1.8pt
                                  \hrule width4.3pt height0.4pt depth0pt}
                             \kern 1pt}
                       {%scriptstyle
                            \kern 0.5pt
               \raise 1pt
                            \vbox{\hrule width4.0pt height0.3pt depth0pt
                                  \kern 1.9pt
                                  \hrule width4.0pt height0.3pt depth0pt}
                            \kern 1pt}
                       {%scriptscriptstyle
                           \kern 0.5pt
             \raise 1pt  \vbox{\hrule width3.6pt height0.3pt depth0pt
                                  \kern 1.5pt
                                  \hrule width3.6pt height0.3pt depth0pt}
                           \kern 0.5pt}
                       }}

\catcode`@=11
\def\un#1{\relax\ifmmode\@@underline#1\else
        $\@@underline{\hbox{#1}}$\relax\fi}
\catcode`@=12

\let\du=\du                     % dot-under
\def\a{\alpha}
\def\b{\beta}
\def\c{\chi}
\def\d{\delta}

\def\f{\phi}
\def\g{\gamma}
\def\h{\eta}

\def\j{\psi}
\def\k{\kappa}
\def\l{\lambda}
\def\m{\mu}
\def\n{\nu}
\def\o{\omega}
\def\p{\pi}
\def\q{\theta}
\def\r{\rho}
\def\s{\sigma}
\def\t{\tau}

\def\x{\xi}
\def\z{\zeta}
\def\D{\Delta}
\def\F{\Phi}
\def\G{\Gamma}

\def\L{\Lambda}
\def\O{\Omega}

\def\S{\Sigma}

\def\ve{\varepsilon}
\def\vf{\varphi}

\def\vq{\vartheta}

\def\cf{{\cal F}}

\def\ck{{\cal K}}
\def\cl{{\cal L}}
\def\cm{{\cal M}}

\def\ct{{\cal T}}

\def\cz{{\cal Z}}

\def\bo{{\raise-.5ex\hbox{\large$\Box$}}}               % D'Alembertian
\def\pa{\partial}                                       % curly d
\def\de{\nabla}                                         % del
\def\pr{\prod}                                          % product
\def\TH{{\raise.2ex\hbox{$\displaystyle \bigodot$}\mskip-4.7mu \llap H \;}}
\def\face{{\raise.2ex\hbox{$\displaystyle \bigodot$}\mskip-2.2mu \llap {$\ddot
        \smile$}}}                                      % happy face
\def\sp#1{{}^{#1}}                              % superscript (unaligned)
\def\slash#1{\rlap{\hbox{$\mskip 1 mu /$}}#1}      % good slash for lower case
\def\Tilde#1{\widetilde{#1}}                    % big tilde
\def\Bar#1{\overline{#1}}                       % big bar
\def\sbar#1{\stackrel{*}{\Bar{#1}}}             % big bar with star

\def\VEV#1{\left\langle #1\right\rangle}        % < >
\def\abs#1{\left| #1\right|}                    % | |
\def\leftrightarrowfill{$\mathsurround=0pt \mathord\leftarrow \mkern-6mu
        \cleaders\hbox{$\mkern-2mu \mathord- \mkern-2mu$}\hfill
        \mkern-6mu \mathord\rightarrow$}
\def\dvec#1{\vbox{\ialign{##\crcr
        \leftrightarrowfill\crcr\noalign{\kern-1pt\nointerlineskip}
        $\hfil\displaystyle{#1}\hfil$\crcr}}}           % <--> accent
\def\dt#1{{\buildrel {\hbox{\LARGE .}} \over {#1}}}     % dot-over for sp/sb
\def\frac#1#2{{\textstyle{#1\over\vphantom2\smash{\raise.20ex
        \hbox{$\scriptstyle{#2}$}}}}}                   % fraction
\def\sfrac#1#2{{\vphantom1\smash{\lower.5ex\hbox{\small$#1$}}\over
        \vphantom1\smash{\raise.4ex\hbox{\small$#2$}}}} % alternate fraction
\def\bfrac#1#2{{\vphantom1\smash{\lower.5ex\hbox{$#1$}}\over
        \vphantom1\smash{\raise.3ex\hbox{$#2$}}}}       % "
\def\afrac#1#2{{\vphantom1\smash{\lower.5ex\hbox{$#1$}}\over#2}}    % "
\def\[{\lfloor{\hskip 0.35pt}\!\!\!\lceil}
\def\]{\rfloor{\hskip 0.35pt}\!\!\!\rceil}
\def\Lag{{\cal L}}
\def\du#1#2{_{#1}{}^{#2}}
\def\ud#1#2{^{#1}{}_{#2}}
\def\dud#1#2#3{_{#1}{}^{#2}{}_{#3}}

\def\fracm#1#2{\hbox{\large{${\frac{{#1}}{{#2}}}$}}}
\def\ha{{\fracmm12}}
\def\tr{{\rm tr}}

\def\un{\underline}
\def\fracmm#1#2{{{#1}\over{#2}}}

\def\low#1{{\raise -3pt\hbox{${\hskip 0.75pt}\!_{#1}$}}}

\def\Dot#1{\buildrel{_{_{\hskip 0.01in}\bullet}}\over{#1}}
\def\dt#1{\Dot{#1}}

\def\Tilde#1{{\widetilde{#1}}\hskip 0.015in}

\newskip\humongous \humongous=0pt plus 1000pt minus 1000pt
\def\caja{\mathsurround=0pt}
\def\eqalign#1{\,\vcenter{\openup2\jot \caja
        \ialign{\strut \hfil$\displaystyle{##}$&$
        \displaystyle{{}##}$\hfil\crcr#1\crcr}}\,}
\newif\ifdtup

\def\ref#1{$\sp{#1)}$}

\def\pl#1#2#3{Phys.~Lett.~{\bf {#1}B} (19{#2}) #3}
\def\np#1#2#3{Nucl.~Phys.~{\bf B{#1}} (19{#2}) #3}
\def\prl#1#2#3{Phys.~Rev.~Lett.~{\bf #1} (19{#2}) #3}
\def\pr#1#2#3{Phys.~Rev.~{\bf D{#1}} (19{#2}) #3}
\def\cqg#1#2#3{Class.~and Quantum Grav.~{\bf {#1}} (19{#2}) #3}
\def\cmp#1#2#3{Commun.~Math.~Phys.~{\bf {#1}} (19{#2}) #3}

\def\ijmp#1#2#3{Int.~J.~Mod.~Phys.~{\bf A{#1}} (19{#2}) #3}

\def\ibid#1#2#3{{\it ibid.}~{\bf {#1}} (19{#2}) #3}

\parskip=\medskipamount                 % space between paragraphs (LaTeX)
\thicklines                         % thick straight lines for pictures (LaTeX)

\begin{document}

% =========================== UH title page ==========================

\thispagestyle{empty}               % no heading or foot on title page (LaTeX)

\def\border{                                            % UH border
        \setlength{\unitlength}{1mm}
        \newcount\xco
        \newcount\yco
        \xco=-24
        \yco=12

        \par\vskip-8mm}

\def\headpic{                                           % UH heading
        \indent
        \setlength{\unitlength}{.8mm}
        \thinlines
        \par
        \par\vskip-6.5mm
        \thicklines}

\border\headpic {\hbox to\hsize{
\vbox{\noindent YITP -- 99 -- 58  \hfill September 1999 \\
ITP--UH--16/99 \hfill hep-th/9909177 \\
DESY 99 -- 142  \hfill revised version }}}

\noindent
\vskip1.3cm
\begin{center}

{\Large\bf Exact low-energy effective actions for }
\vglue.1in
{\Large\bf  hypermultiplets in four dimensions}

\vglue.3in

Sergei V. Ketov \footnote{
Also at High Current Electronics Institute of the Russian Academy of Sciences,
Siberian Branch, \newline ${~~~~~}$ Akademichesky~4, Tomsk 634055, Russia}

{\it Yukawa Institute for Theoretical Physics }\\
{\it Kyoto University, Kyoto 606--8502, Japan}\\
{\sl ketov@yukawa.kyoto-u.ac.jp}

and

{\it Institut f\"ur Theoretische Physik, Universit\"at Hannover}\\
{\it Appelstra\ss{}e 2, Hannover 30167, Germany}\\
{\sl ketov@itp.uni-hannover.de}
\end{center}
\vglue.2in
\begin{center}
{\Large\bf Abstract}
\end{center}

\noindent
We consider the general hypermultiplet {\it Low-Energy Effective Action}
(LEEA) that may appear in quantized, four-dimensional, N=2 supersymmetric, 
gauge theories, e.g. in the Coulomb and Higgs branches. Our main 
purpose is a description of the {\it exact} LEEA of $n$ magnetically 
charged hypermultiplets. The hypermultiplet LEEA is given by the N=2 
supersymmetric {\it Non-Linear Sigma-Model} (NLSM) with a $4n$-dimensional 
hyper-K\"ahler metric, subject to non-anomalous symmetries. {\it Harmonic 
 Superspace} (HSS) and the NLSM isometries are very useful to constrain the 
hyper-K\"ahler geometry of the LEEA. We use N=2 supersymmetric projections of 
HSS superfields to N=2 linear (tensor) $O(2)$ and $O(4)$ multiplets in N=2 
{\it Projective Superspace} (PSS) to deduce the explicit form of the LEEA in 
some particular cases. As the by-product, a simple new classification of all 
multi-monopole moduli space metrics having $su(2)_R$ symmetry is proposed in 
terms of real quartic polynomials of $2n$ variables, modulo $Sp(n)$ 
transformations. The 4d hypermultiplet LEEA for $n=2$ can be encoded in terms 
of an elliptic curve. 

\newpage

\section{Introduction}

The seminal work of Seiberg and Witten \cite{sw} gave many important
insights into the non-perturbative dynamics of the four-dimensional
N=2 supersymmetric $SU(2)$ Yang-Mills theory whose gauge symmetry is 
spontaneously broken to its abelian subgroup (in the so-called Coulomb branch).
It was subsequently generalized to other simply-laced gauge groups and 
N=2 super-QCD as well \cite{swg} (see also ref.~\cite{rev} for 
a review or an introduction). Seiberg-Witten theory deals with the exact 
{\it Low-Energy Effective Action} (LEEA) in terms of {\it abelian} N=2
vector multiplets, which includes both perturbative (one-loop) and 
non-perturbative (instanton) corrections. To fix those corrections, N=2 
extended supersymmetry plays the important r\^ole that amounts to restricting
the N=2 vector gauge LEEA to the general `Ansatz' governed by a single 
{\it holomorphic} function $\cf(W^A)$ of the N=2 abelian vector superfield 
strengths $W^A$, where $A=1,2,\ldots,N_c-1$  \cite{poten,poten2}. This Ansatz 
is manifestly N=2 supersymmetric, gauge-invariant  and model-independent due 
to the off-shell nature of the N=2 restricted chiral superfields  $W^A$. 
According to the Riemann-Hilbert theorem, a (milti-valued) holomorphic 
function is fully determined by its singularity structure and monodromy 
(or asymptotics). The number of singularities is dictated by Witten index 
\cite{ind}. However, a calculation of the Witten index in massless gauge field
theories is plagued with ambiguities, so that its value is usually postulated 
from physical considerations and consistency. This is closely related to the
existence of BPS monopoles representing the non-perturbative degrees of 
freedom in the non-abelian N=2 gauge theory under consideration \cite{rev}. 
Indeed, any non-abelian N=2 supersymmetric gauge field theory can be 
considered as a particular Yang-Mills-Higgs system (with adjoint fermions) 
whose field equations admit the solitonic solutions labelled by magnetic 
charge, and whose scalar potential is fixed by N=2 supersymmetry.
The monodromy behaviour is obtained from perturbative calculations and 
electric-magnetic (S) duality. For example, the chiral anomaly of the
underlying N=2 gauge theory fully determines the perturbative (logarithmic) 
contribution to the second derivative of $\cf(W)$, which is equivalent
to knowing the monodromy around the infinity in the quantum moduli space 
parametrized by vacuum expectation values of the Higgs fields. The S
duality and global consistency conditions fix the remaining
monodromies and, hence, the whole function $\cf(W)$. 

Due to the holomorphicity of the Seiberg-Witten LEEA, the
non-perturbative contributions to $\cf(W)$ take the form of an
infinite sum over all instanton numbers. In 
particular, no instanton/anti-instanton (i.e. non-holomorphic) contributions
can appear in the Seiberg-Witten LEEA, in contrast to the standard (N=0) QCD.
Yet another special and very remarkable feature of the Seiberg-Witten-type 
solutions is the fact that any of them (e.g., in the $SU(N_c)$-based theory)
can be nicely encoded in terms
of the auxiliary (hyperelliptic) curve (or Riemann surface) $\S_{\rm SW}$
of genus $(N_c-1)$ \cite{sw,swg}. The function $\cf(W)$ is most 
naturally represented in the parametric form, in terms of certain abelian
differential $\l_{\rm SW}$ (of the 3rd kind) integrated over the periods 
of $\S_{\rm SW}$ \cite{rev}.

N=2 supersymmetric gauge field theories can have only two types of rigid N=2 
supermultiplets, an N=2 {\it vector} multiplet comprising the helicities 
$(\pm 1,\pm\ha,0)$, and a {\it hypermultiplet} comprising the helicities 
$(\pm\ha,\pm 0)$. The two types of N=2 multiplets are truly  different
in {\it four} spacetime dimensions (4d), but they become dual to each other in 
{\it three} dimensions (3d). The latter observation can be elevated
 to the co-called {\it `c-map'} relating the special K\"ahler geometry
of the vector multiplet moduli space to the hyper-K\"ahler geometry
of the hypermultiplet moduli space \cite{cfg} or, even further, to the 
mirror symmetry between certain type-II superstring compactifications on
Calabi-Yau manifolds \cite{polch}. The LEEA of the 3d, N=4 supersymmetric
non-abelian gauge field theories can, therefore, be deduced along the
lines of the 4d Seiberg-Witten theory \cite{sw3}. One of the most
remarkable developments in the 3d, N=4 Seiberg-Witten theory was the
proposed equivalence between the {\it quantum} moduli space of the 3d, N=4 
supersymmetric pure $SU(n)$ gauge theory (in the Coulomb branch) and the {\it 
classical} moduli space of $n$ BPS monopoles in the 4d, $SU(2)$-based 
Yang-Mills-Higgs system \cite{ch,hwit}. In particular, as was argued by 
Hanany and Witten \cite{hwit}, this equivalence between the moduli spaces 
of {\it different} field theories in {\it different} dimensions, for all 
$n$, can be understood as a consequence of S-duality applied to certain 
configurations of intersecting (Dirichlet) 3-branes and 5-branes in type-IIB 
superstring theory. Though the brane technology is very efficient in 
explaining the equivalence between the apparently different moduli
spaces, it is not powerful enough in deriving explicit metrics on them. In 
this paper we consider those metrics possessing an $su(2)_R$ isometry. Another
 class of the $so(2)$-invariant hyper-K\"ahler metrics arises in the quantum 
moduli spaces of the 3d, N=4 supersymmetric $SU(2)$ gauge theories with $k$ 
fundamental matter hypermultiplets in the Coulomb branch. The latter can be 
identified with $D_k$ gravitational instantons in four Euclidean dimensions. 
The $D_k$ metrics in (almost) explicit form were recently calculated in 
ref.~\cite{chka} from the standard Nahm construction \cite{adhmn}, see also 
refs.~\cite{dancer,chal2} for the related work. In the purely gauge 3d, 
N=4 theory with $N_c=2$ the quantum moduli space metric is known to be given 
by the $su(2)_R$ invariant Atiyah-Hitchin metric \cite{sw3,dkmtv}. The 
{\it asymptotical} metrics in the classical multi-monopole moduli spaces for 
well-separated monopoles in 4d are available in their explicit form 
\cite{lwp,gr}, whereas the {\it exact} metrics are only known up to certain 
algebraic (Ercolani-Sinha) constraints \cite{ersin,hmr}.

A derivation of the exact hypermultiplet LEEA in four spacetime dimensions (4d)
requires the techniques that are very different from the ones used in the 
Seiberg-Witten theory. The main reason is the different status of an N=2 
scalar multiplet (called a {\it hypermultiplet}) versus an N=2 vector 
multiplet. To appreciate this fact, it is worth mentioning the well-known fact
that there exist {\it no} off-shell, manifestly N=2 supersymmetric  
(i.e. model-independent) formulation of the most fundamental 
{\it Fayet-Sohnius} (FS) hypermultiplet in the conventional N=2 extended 
superspace (see, e.g., ref.~\cite{oldket} for a review of N=2 superfields, and
 the references therein). Some restricted (non-universal) off-shell versions 
of a hypermutiplet, nevertheless, exist in N=2 {\it Projective Superspace} 
(PSS) invented by Karlhede, Lindstr\"om and Ro\v{c}ek \cite{klr}, where they 
are known as projective \cite{klr,hklr} or (generalised) tensor N=2 mutliplets 
\cite{ok}. The PSS construction gives up the manifest $su(2)_R$ internal 
symmetry rotating N=2 supersymmetry charges, while it also implies vector 
fields amongst the N=2 projective superfield components. The most symmetric 
approach to N=2 supersymmetry is provided by {\it Harmonic Superspace} (HSS) 
invented by Galperin, Ivanov, Kalitzin, Ogievetsky and Sokatchev \cite{gikos},
by using the infinite number of auxiliary fields. Unlike the PSS approach, 
both N=2 vector multiplets {\it and} hypermultiplets can be introduced in HSS 
{\it on equal footing}. Moreover, HSS allows one to keep both N=2 
supersymmetry {\it and} its $su(2)_R$ automorphisms manifest. In the PSS 
approach, one adds a holomorphic (projective) coordinate to the conventional 
N=2 superspace, and then one uses hidden holomorphicity of the 
N=2 superspace constraints defining the N=2 projective  multiplets. 
In the HSS description, one adds group-valued twistors and uses hidden 
Grassmann analyticity of the off-shell HSS constraints defining 
N=2 vector multiplets and hypermultiplets. In our view, as far as 
N=2 supersymmetry in 4d is concerned, the Grassmann analyticity is 
more fundamental than holomorphicity. The PSS and HSS descriptions of the 
hypermultiplet LEEA are closely related, most notably, by the N=2 
supersymmetric projections of HSS superfields onto the PSS superfields.
  
The group of analyticity-preserving field reparametrizations (in HSS) is much 
larger than the group of holomorphicity-preserving reparametrizations 
(in PSS). Accordingly, a derivation of the exact hypermultiplet LEEA is more 
complicated or, at least, different, from solving a Seiberg-Witten 
(or Riemann-Hilbert) problem, just because more data is needed to fix a 
(Grassmann) analytic function versus a holomorphic one. A formal solution to 
the Riemann-Hilbert problem is given by the linear system of Picard-Fuchs 
differential equations \cite{fre}. As is demonstrated in this paper, a formal 
solution to the hypermultiplet LEEA can be given in HSS, in the form of an 
analytic hyper-K\"ahler potential of the metric. A derivation of the metric 
from the hyper-K\"ahler potential requires an elimination of all the auxiliary
 fields hidden in analytic hypermultiplet superfields, which amounts to 
solving the (infinite) linear system of differential equations on a two-sphere.

From the physical point of view, it is important to understand the
origin of a non-trivial hypermultiplet self-interaction in the LEEA. 
For example, as was argued by Seiberg and Witten \cite{sw}, the exact 
effective NLSM metric for `fundamental' hypermultiplets with vanishing 
magnetic charges in the Higgs branch of a 4d, N=2 gauge field theory 
is {\it flat}, i.e. there is no self-interaction at all. However, 
{\it magnetically} charged (massive) hypermultiplets can have a 
non-trivial self-interaction \cite{hk1,ikz}. This observation is 
consistent with the brane technology  \cite{rev2}. The corresponding
LEEA just describes the low-energy dynamics of the BPS monopoles
representing nonperturbative degrees of freedom, in the
Lorentz-invariant way ({\it cf.} ref.~\cite{manton}). For instance, 
the non-trivial NLSM corrections to the perturbative hypermultilet
LEEA found in refs.~\cite{hk1,ikz} (see sect.~3 too) are all
proportional to the squared absolute value of a central charge in
N=2 supersymmetry algebra. The central charge itself is given by a
linear combination of abelian charges of the underlying 
(spontaneously broken) non-abelian N=2 gauge theory. The non-vanishing 
central charge is also responsible for a dynamical generation of the 
non-trivial {\it scalar} potential associated with the hypermultiplet
LEEA \cite{ikz,ku}. Those important features were not fully appreciated in 
the earlier investigations of quantized 4d, N=2 supersymmetric field theories,
which were either limited to renormalizable N=2 field theories or didn't 
include N=2 central charges into the propagators (see, however, 
ref.~\cite{poten2} where the scalar potentials based on `active' central 
charges were investigated in 2d, N=4 NLSM). 

The paper is organized as follows. In sect.~2 we review a superspace
derivation of the known relation between N=2 supersymmetry and hyper-K\"ahler 
geometry in 4d NLSM, which plays the important r\^ole in our investigation. We
remind the reader about the formulation of N=2 NLSM in N=1 superspace and then
motivate the use of PSS and HSS for a resolution of the hyper-K\"ahler 
constraints and isometries. In sect.~3 we discuss in detail the most
general 4d LEEA of a single hypermultiplet, and demonstrate that it is
given by the N=2 NLSM with the Taub-NUT or Atiyah-Hitchin metric. Sect.~4 is
devoted to a discussion of the LEEA for many hypermultiplets and related
multi-monopole moduli space metrics. Sect.~5 comprises our conclusion. Our 
presentation is self-contained. Basic facts about hyper-K\"ahler geometry are 
collected in Appendix A. Projective superspace (PSS) is introduced in Appendix
 B. Harmonic superspace (HSS) is defined in Appendix C. The classical moduli 
spaces of solitonic solutions in the 4d, $SU(2)$-based Yang-Mills-Higgs system
 are briefly reviewed in Appendix D.

\section{Supersymmetry and hyper-K\"ahler geometry}

The most natural description of 4d supersymmetry is provided by
superspace. The natural framework for N=1 supersymmetry is
given by N=1 superspace \cite{sspace}. N=2 extended supersymmetry
can be manifestly realized in N=2 superspace that has three different 
versions (standard, projective and harmonic). As regards general N=2 NLSM,
in subsects.~2.1 and 2.2 we first recall their description in N=1 superspace 
\cite{haklr} and then in N=2 HSS \cite{gios,gios2}, in order to remind the 
reader about the equivalence between rigid N=2 supersymmetry and 
hyper-K\"ahler geometry. Isometries in general N=2 NLSM are discussed in 
subsect.~2.3. The whole section establishes our setup and provides a technical 
introduction to the rest of the paper.

\subsection{N=2 NLSM in N=1 superspace}

N=1 scalar (chiral) multiplets are described by the N=1 
complex chiral superfields $\F^i$ and their conjugates $\bar{\F}_i$, 
$i=1,2,\ldots,k$, satisfying the off-shell constraints
$$ \bar{D}_{\dt{\a}}\F^i=0~,\quad D\low{\a}\bar{\F}_i=0~,\eqno(2.1)$$
where we have introduced the covariant spinor derivatives 
$D\low{\a}$ and $\bar{D}_{\dt{\a}}$ in flat N=1 superspace 
$\cz=(x^{\m},\q^{\a},\bar{\q}^{\dt{\a}})$. They obey the basic 
anticommutation relations of N=1 supersymmetry,
$$ \{ D\low{\a},\bar{D}_{\dt{\a}}\}=i\pa_{\a\dt{\a}}~,\quad
\{D\low{\a},D\low{\b}\}=\{\bar{D}_{\dt{\a}},\bar{D}_{\dt{\b}} \}=0~.
\eqno(2.2)$$
We use the two-component spinor notation that is standard in 4d supersymmetry 
\cite{sspace}. The field components of the chiral superfield $\F^i$ are 
$$ A^i=\left.\F^i\right|~,\quad \j^i_{\a}=D_{\a}\left.\F^i\right|~,\quad 
F^i=\frac{1}{2}D^{\a}D_{\a}\left.\F^i\right|~,\eqno(2.3)$$
where $\left.\right|$ means taking the first (leading, or $\q$-independent) 
component of a superfield or an operator. The scalars $A$ and the
spinors $\j$ are the propagating fields, whereas the scalars $F$ are
the auxiliary fields. 

The general 4d NLSM is described by an action
$$ S_{\rm NLSM}[A] = \fracmm{1}{2\k^2}\int d^4x\,g_{ij}(A)\pa_{\m}A^i
\pa^{\m}A^j~.\eqno(2.4)$$
It has an N=1 supersymmetric extension if and only if the NLSM metric 
$g_{ij}(A)$ is {\it K\"ahler} \cite{zumino}. Indeed, the most  general 
N=1 supersymmetric action, in terms of N=1 chiral superfields and of 
the second order in spacetime derivatives of the physical scalars, reads
$$ S = \int d^4x d^4\q\, K(\F^i,\bar{\F}_j)=-\frac{1}{2}\int d^4x\,
\left.K\du{i}{j}\right|\pa^{\a\dt{\a}}A^i\pa_{\a\dt{\a}}\bar{A}_j+\ldots~,
\eqno(2.5)$$
where we have explicitly written down the leading bosonic NLSM term. 
We use the notation \cite{haklr}
$$ K_{i_i\cdots i_n}^{j_1\cdots j_m} \equiv \fracmm{\pa}{\pa A^{i_1}}\cdots 
\fracmm{\pa}{\pa A^{i_n}} \fracmm{\pa}{\pa \bar{A}_{j_1}}\cdots
\fracmm{\pa}{\pa \bar{A}_{j_m}} K(A,\bar{A})~,\eqno(2.6)$$
and similarly for $K(\F^i,\bar{\F}_j)$. The right-hand-side 
of eq.~(2.5) has the standard NLSM form (2.4) with the restricted
(=K\"ahler) metric
$$ ds^2=K\du{i}{j}(A,\bar{A}) dA^id\bar{A}_j~.\eqno(2.7)$$
A complex manifold whose metric can be written down in the form (2.7) with
a locally defined potential $K$ is called the {\it K\"ahler} manifold
\cite{difg}. The form of the superfield NLSM action (2.5) is preserved 
under arbitrary reparametrizations of $\F^i$ {\it and} $\bar{\F}_j$.
The line element (2.7) is only preserved under {\it holomorphic}
 reparametrizations of $A^i$ and $\bar{A}_j$, $A^i\to f^i(A^j)$, while
this can be extended to holomorphic transformations of the chiral superfields,
$\F^i\to f^i(\F^j)$. The Ricci tensor of a K\"ahler metric reads
$$ R\du{i}{j}=\{\ln\det(K\du{k}{m})\}\du{,i}{,j}~~. \eqno(2.8)$$

Having constructed the most general N=1 supersymmetric NLSM action
(2.5) in terms of a K\"ahler potential $K(\F,\bar{\F})$, one can further
impose extra non-manifest (non-linear) supersymmetry on the action
(2.5), in order to get N=2 NLSM. In the absence of N=2 auxiliary fields, the
extended N=2 supersymmetry algebra can be closed only on-shell, i.e. on
the equations of motion for the NLSM fields.
 
The most general `Ansatz' for the transformation law of extra
supersymmetry is given by 
$$ \d\F^i=\bar{D}^2(\bar{\ve}\bar{\O}^i)~,\quad 
\d\bar{\F}_i=D^2(\ve \O_i)~,\eqno(2.9)$$
where $\ve$ is a constant chiral superfield parameter,
$\bar{D}_{\dt{\a}}\ve=D^2\ve=\pa_{\m}\ve=0$, and $\bar{\O}$ is a
function of $\F$ {\it and} $\bar{\F}$ (modulo an additive chiral
term). The on-shell closure of the supersymmetry transformations
(2.9) implies the relations  \cite{haklr}
$$ \eqalign{
\O_{i,j}\bar{\O}^{j,k}=& \O_{j,i}\bar{\O}^{k,j}=  -\d^k_i~,\cr
\bar{\O}^{j,[m|}\bar{\O}\ud{i,|k]}{j}= & ~0~,\cr
\bar{D}^2\bar{\O}^i=& \bar{\O}^{i,j}\bar{D}^2\bar{\F}_j+
\frac{1}{2}\bar{\O}^{i,jk}\bar{D}_{\dt{\a}}\bar{\F}_j
\bar{D}^{\dt{\a}}\bar{\F}_k=0~,\cr}\eqno(2.10)$$
and their complex conjugates. The N=1 NLSM action (2.5) is invariant 
under the transformations (2.9) provided that \cite{haklr}
$$\eqalign{
\bar{\o}^{jm}\equiv K\du{i}{j}\O^{i,m}=& -\bar{\o}^{mj}~,\cr
K\du{i}{j}\bar{\O}^{i,mk}+K\du{i}{mk}\bar{\O}^{i,j}=& ~0~,\cr
K\du{i}{j}\bar{\O}\ud{i,m}{k}+K\du{ik}{j}\bar{\O}^{i,m}=& ~0~.\cr}
\eqno(2.11)$$
This is to be compared to the field equations following from the action (2.5),
$$ \bar{D}^2K_i=K\du{i}{j}\bar{D}^2\bar{\F}_j +
\frac{1}{2}K\du{i}{jk}\bar{D}_{\dt{\a}}\bar{\F}_j
\bar{D}^{\dt{\a}}\bar{\F}_k=0~.\eqno(2.12)$$
The first two lines of eq.~(2.11) imply that the third line of
eq.~(2.10) is equivalent to the equation of motion (2.12), which
confirms the on-shell closure of the N=2 supersymmetry algebra \cite{haklr}.

It is now straightforward to check that eqs.~(2.10) and (2.11)
together amount to hyper-K\"ahler geometry (Appendix A). In
particular, the quaternionic structure (A.12) comprises
the canonical complex structure $J^{(3)}$ of eq.~(A.6)  
and two non-canonical complex structures, 
$$ J^{(1)j}_{i}=\left(\begin{array}{cc} 0 & \O_{j,i} \\
\bar{\O}^{j,i} & 0  \end{array} \right) \quad{\rm and}\quad
J^{(2)j}_{i}=\left(\begin{array}{cc} 0 & i\O_{j,i} \\
-i\bar{\O}^{j,i} & 0 \end{array} \right)~,\eqno(2.13)$$
with mixed (one covariant and one contravariant) indices ---  see 
the first line of eq.~(2.10). Both $J^{(1)}$ and $J^{(2)}$ 
are integrable due to the second line of eq.~(2.10), while they are 
covariantly constant due to the second and third lines of eq.~(2.11). 
Finally, the NLSM metric is hermitian with respect to all three
 complex structures due to the first line of eq.~(2.11). According 
to Appendix A, this precisely amounts to the hyper-K\"ahler structure. 

The canonical complex structure, $J^{(3)}$, is obviously related
to the given K\"ahler structure of the N=1 NLSM that we started
with. The coordinate system, where the metric takes the
K\"ahler form with respect to a non-canonical complex structure, 
is, therefore, related to the preferred one by a {\it nonholomorphic}
coordinate transformation.

Eq.~(2.13) gives the complex structures in terms of the 
derivatives of the nonholomorphic functions $\O_i$ and
$\bar{\O}^i$ introduced in eq.~(2.9). These functions
can be reconstructed from a given K\"ahler potential and one 
of the noncanonical complex structures $J$ \cite{haklr}. Since 
$J$ anticommutes with the canonical complex structure, the
former can be written down in the form
$$ J\du{i}{j} = \left( \begin{array}{cc} 0 & \O_{ji} \\
\bar{\O}^{ji} & 0 \end{array} \right) \eqno(2.14)$$
with some matrix $\O$ and its complex conjugate
$\bar{\O}$. It implies that $J^{ij}$ is block-diagonal,
$$ J^{ij} =\left( \begin{array}{cc} \bar{\g}^{ij} & 0 \\
0 & \g_{ij}  \end{array} \right)~, \quad {\rm where}\quad 
\bar{\g}^{ij}\equiv (K^{-1})^i_k\bar{\O}^{kj}~.\eqno(2.15)$$
The covariant constancy of $J$ yields that
$\bar{\g}^{jk}$ is holomorphic whereas $\g_{jk}$ is
antiholomorphic, $\bar{\pa}^i\bar{\g}^{jk}=\pa_i\g_{jk}=0$.
Similarly one finds that $\bar{\o}^{ij}$, introduced in the
first line of eq.~(2.11), is antiholomorphic, whereas $\o_{ij}\equiv
K\du{i}{k}\O_{kj}$ is holomorphic.

It is straightforward to verify that the functions \cite{haklr}
$$\bar{\O}^i\equiv \bar{\g}^{ij}K_j \eqno(2.16)$$
obey the desired relation
$$  \bar{\O}^{i,j} = \bar{\O}^{ij}~,\eqno(2.17)$$
while they satisfy all eqs.~(2.10) valid for any hyper-K\"ahler
manifold. Eqs.~(2.15) and (2.16) further imply that
$$ K_i\bar{\O}^i=0 \quad {\rm and}\quad \de_i\bar{\O}^i=0~.
\eqno(2.18)$$
One concludes that the hyper-K\"ahler structure plays the most 
fundamental role in the hyper-K\"ahler geometry, since all other
geometrical quantities can be constructed in terms of it.

The N=1 superspace approach remains the most popular method in 4d 
supersymmetry, mainly because it has a very simple and clear
connection to the component approach. However, N=1 superspace is
clearly inadequate for N=2 supersymmetry since only one of the 
supersymmetries can be manifestly realized there, whereas another
 supersymmetry is necessarily hidden, being non-linearly realized.
The differential constraints implied by the second 
supersymmetry (see, e.g., eq.~(A.13) in Appendix A) are to be solved,
 since their presence does not allow one to formulate the most general
`Ansatz` for the N=2 hypermultiplet LEEA or, equivalently, the most general 
hyper-K\"ahler NLSM. Perhaps, most importantly, any treatment of 
{\it isometries} of N=2 NLSM is very complicated in N=1 superspace 
\cite{haklr}. The N=2 NLSM isometries are, however, going to be crucial for 
our purposes in the next sections.

\subsection{N=2 superspace and NLSM}

The 4d, N=2 superalgebra $SUSY^2_4$ is a graded extension of the 4d
Poincar\'e algebra. In addition to the generators of the Poincar\'e algebra
$(P_{\m}, M_{\l\r})$, $\m=0,1,2,3$, the superalgebra $SUSY^2_4$
contains two Majorana  spinor generators $(Q^i_{\a},\bar{Q}^{\dt{\a}}_i)$,
$i=1,2,$ and the generators $A\ud{i}{j}$ of the $U(2)_R=SU(2)_R\times U(1)_R$
automorphisms. Together they satisfy the (anti)commutation relations
$$ \frac{1}{i} \[ M_{\m\n},M_{\r\l} \]~=~\h_{\n\r}M_{\m\l}-\h_{\n\l}M_{\m\r}
+\h_{\m\l}M_{\n\r}-\h_{\m\r}M_{\n\l}~,$$
$$ \frac{1}{i} \[ P_{\m},M_{\n\l} \]~=~\h_{\m\n}P_{\l}-\h_{\m\l}P_{\n}~,\quad
\[P_{\m},P_{\n}\]~=~0~,$$
$$\[Q^i_{\a},M_{\m\n}\]~=~\frac{i}{4}(\s_{\m\n}Q^i)_{\a}~,\quad 
\[\bar{Q}_j^{\dt{\a}},M_{\m\n}\]~=~\frac{i}{4}
(\tilde{\s}_{\m\n}\bar{Q}_j)^{\dt{\a}}~,$$
$$
\[P_{\m},Q^i_{\a}\]~=~\[P_{\m},\bar{Q}^{\dt{\a}}_i\]~=~\{Q^i_{\a},Q^j_{\b}\}
~=~\{\bar{Q}^{\dt{\a}}_i,\bar{Q}^{\dt{\b}}_j\}~=~0~,$$
$$
\{Q^i_{\a},\bar{Q}_{j\dt{\a}}\}~=~\d\ud{i}{j}\s^{\m}_{\a\dt{\a}}P_{\m}~,$$
$$\[A\ud{i}{j},A\ud{l}{m}\]~=~\d\ud{l}{j}A\ud{i}{m}-\d\ud{i}{m}A\ud{l}{j}~,$$
$$ \[A\ud{i}{j},Q^l_{\a}\]~=~\d\ud{l}{j}Q^i_{\a},~\quad
\[A\ud{i}{j},\bar{Q}^{\dt{\a}}_l\]~=~-\d\ud{i}{l}\bar{Q}^{\dt{\a}}_j~,$$
$$\[A\ud{i}{j},M_{\m\n}\]~=~\[A\ud{i}{j},P_{\m}\]~=~0~.\eqno(2.19)$$
The $U(1)_R$ generator $B\equiv A\ud{i}{i}$ has the commutation relations
$$ \[B,Q^i_{\a}\]=Q^i_{\a}~,\qquad  \[B,\bar{Q}^{\dt{\a}}_i\]=
-\bar{Q}^{\dt{\a}}_i~.\eqno(2.20)$$

In flat N=2 superspace with the coordinates 
$\cz=(x^{\m},\q^{\a}_i,\bar{\q}^i_{\dt{\a}})$ one can also introduce the N=2 
covariant spinor derivatives $(D^i_{\a},\bar{D}_j^{\dt{\a}})$ anticommuting
with the N=2 supersymmetry charges and satisfying the same algebra,
$$ \{D^i_{\a},\bar{D}_{j\dt{\a}}\}=\d\ud{i}{j}\s^{\m}_{\a\dt{\a}}P_{\m}~,\quad
 \{D^i_{\a},D^j_{\b}\}=\{\bar{D}^{\dt{\a}}_i,\bar{D}^{\dt{\b}}_j\}=0~.
\eqno(2.21)$$
Their explicit realization reads
$$ D^i_{\a}=\fracmm{\pa}{\pa\q^{\a}_i}-\frac{i}{2}\pa_{\a\dt{\a}}
\bar{\q}^{\dt{\a}i}~,\quad 
\bar{D}^{\dt{\a}}_j=\fracmm{\pa}{\pa\bar{\q}^j_{\dt{\a}}}-\frac{i}{2}
\pa^{\dt{\a}\a}\q_{\a j}~.\eqno(2.22)$$

A general N=2 superfield is the {\it irreducible} representation of the
{\it enlarged} superalgebra $SUSY^{2,D}_4$ defined by adding the
covariant derivatives $(D^i_{\a},\bar{D}_j^{\dt{\a}})$ to the generators of
$SUSY^2_4$. The same superfield is, however, {\it reducible} with 
respect to the N=2 supersymmetry algebra $SUSY^2_4$, while its irreducible 
constituents can be defined either by imposing certain N=2 superspace 
constraints or by using superprojectors \cite{sig,oldket}. For example, the 
abelain N=2 superfield strength of an N=2 vector multipet is desribed by the 
restricted chiral N=2 superfield $W$ subject to the N=2 superspace (off-shell)
 constraints \cite{wess,oldket}
$$\bar{D}^{\dt{\a}}_i W=0~,\quad  D^4W=\bo \bar{W}~,\quad{\rm where}\quad
D^4\equiv \prod_{i,\a}^4 D^i_{\a}~.\eqno(2.23)$$
Solving the constraints (2.23) is fully straightforward, and it
results in \cite{wess,oldket}
$$\eqalign{
W=\exp\left\{-\frac{i}{2}\q_i\slash{\pa}\bar{\q}^i\right\} & \left[
A + \q^{\a}_i\j^i_{\a}-\ha\q^{\a}_i(\t_m)\ud{i}{j}\q^j_{\a}C_m
+\frac{1}{8}\q^{\a}_i(\s_{\m\n})\du{\a}{\b}\q^i_{\b}F^{\m\n}\right.\cr
& \left. -i(\q^3)^{i\a}\pa_{\a\dt{\b}}\bar{\j}^{\dt{\b}}_i
+\q^4\bo\bar{A}\right]~,\cr}\eqno(2.24)$$
in terms of the complex scalar $A$, the Majorana douplet $\j^i_{\a}$, the real
auxiliary $SU(2)$ triplet $C_m$, and the real antisymmetric tensor
$F_{\m\n}$  subject to the `Bianchi identity',
$$ \ve^{\m\n\l\r}\pa_{\n}F_{\l\r}=0~, \eqno(2.25)$$
whose solution is $F_{\m\n}=\pa_{\m}A_{\n}-\pa_{\n}A_{\m}$.

The free equations of motion for the N=2 vector multiplet in the
standard N=2 superspace are given by
$$ D_{ij}W=0~,\quad {\rm where}\quad D_{ij}\equiv D^{\a}_iD_{\a j}~.
\eqno(2.26)$$

Let's now introduce the N=2 source superfield $L_{ij}=L_{ji}$ into eq.~(2.26),
$$ D_{ij}W=L_{ij}~.\eqno(2.27)$$
The constraints (2.23) imply the constraints on the N=2 superfield $L_{ij}$,
$$ D^{(i}_{\a}L^{jk)}=\Bar{D}_{\dt{\a}}{}^{(i}L^{jk)}=0 \quad {\rm and}\quad
L^{ij}=\ve^{ik}\ve^{jl}\Bar{L}_{kl}~.\eqno(2.28)$$
The constraints (2.28) define an off-shell N=2 {\it tensor} multiplet in
the standard N=2 superspace \cite{klr}.

Eq.~(2.28) can be further generalized to N=2 superfields  $L^{i_1\cdots i_n}$ 
that are totally symmetric with respect to their 
$SU(2)_{\rm R}$ indices and satisfy the constraints \cite{ok}
$$ D_{\a}^{(k}L^{i_1\cdots i_n)}=\Bar{D}_{\dt{\a}}{}^{(k}L^{i_1\cdots i_n)}
=0~.\eqno(2.29)$$
In the case of an even number of indices, $n=2p$, the superfields 
$L^{i_1\cdots i_{2p}}$ can satisfy the reality condition
$$ \Bar{L}_{i_1\cdots i_{2p}}\equiv (L^{i_1\cdots i_{2p}})^*=\ve_{i_1j_1}\cdots
\ve_{i_{2p}j_{2p}}L^{i_1\cdots i_{2p}}~~.\eqno(2.30)$$

Eqs.~(2.29) and (2.30) define the {\it projective} (generalized
tensor) N=2 multiplets for all $n\geq 2$ \cite{klr,ok}. They are
irreducible off-shell representations of N=2 extended 4d supersymmetry
of superspin $Y=0$ and superisospin $I=(n-2)/2$, having 
$8(n-1)$ bosonic and the same number of fermionic (off-shell) field components 
\cite{ok},
$$ \left\{ L^{i_1\cdots i_{n}}~,\quad \j_{\a}^{i_2\cdots
i_n}~,\quad  N^{i_3\cdots i_n}~,\quad 
V_{\a\dt{\a}}^{i_3\cdots i_n}~,\quad \bar{\j}_{\dt{\a}}^{i_4\cdots
i_n}~,\quad C^{i_5\cdots i_n}\right\}~.\eqno(2.31)$$
It follows from matching the bosonic and fermionic degrees of freedom 
that the real vector $V_{\a\dt{\a}}$ in an N=2 tensor multiplet has to be
{\it conserved}, $\pa^{\a\dt{\a}}V_{\a\dt{\a}}=0$, whereas the vector fields 
$V_{\a\dt{\a}}^{i_3\cdots i_n}$ of the projective N=2 multiplets with 
$n>2$ are all unconstrained.  

Choosing $n=1$ in eq.~(2.29) results in the {\it on-shell} constraints 
defining the {\it Fayet-Sohnius} (FS) hypermultiplet (of vanishing central 
charge) whose independent components have helicities $(\pm\ha,\pm 0)$, 
as is required for the `true' hypermultiplet \cite{fas}. The failure to 
incorporate the off-shell (i.e. model-independent) FS hypermultiplet within 
the framework of the standard N=2 superspace has far reaching consequences 
in N=2 supersymmetry. In particular, as regards N=2 NLSM, it does not allow 
one to formulate the most general, manifestly N=2 supersymmetric 
NLSM in the standard N=2 superspace. Since N=2 supersymmetry amounts to
the hyper-K\"ahler NLSM geometry, the roots of the problem can be traced
back to the basic properties of the hyper-K\"ahler structure (Appendix A).
To the end of this subsection, we argue that the {\it twistor} approach 
 \cite{gios,gios2} is indeed the natural way to solve this problem.

Given a hyper-K\"ahler manifold $\cm$, a linear combination,
$aJ^{(1)}+bJ^{(2)}+cJ^{(3)}$,  of its three, linearly independent and 
covariantly constant, complex structures satisfying the quaternionic 
algebra (A.12), with arbitrary real parameters $(a,b,c)$, is also the
covariantly constant complex srtructure provided that $a^2+b^2+c^2=1$.
Hence, a hyper-K\"ahler manifold $\cm$ possess the variety of non-canonical
complex structures (on the top of the canonical one), worthy of a
two-sphere $S^2$. This feature is crucial for the efficiency of the twistor 
space \cite{ati} in monopole physics, whereas its PSS and HSS extensions 
provide the natural framework for an explicit construction of hyper-K\"ahler 
metrics from superspace.
In the HSS approach one extends the ordinary N=2 superspace by the two-sphere
 $S^2$. Because of the isomorphism $S^2\sim SU(2)/U(1)$, one can actually add
 the group $SU(2)$ instead of the coset, by restricting the HSS superfields 
to the ones that are {\it equivariant} with respect to the $U(1)$ symmetry  
--- this mathematical construction is the very particular realization of a 
{\it flag manifold} \cite{flag}.  The $SU(2)$ symmetry can be identified with 
the N=2 supersymmetry automorphisms $SU(2)_R$ that can be made manifest in 
HSS. In the PSS construction \cite{klr,hklr} one uses another isomorphism 
$S^2\sim CP(1)$ by adding a (complex) projective line $CP(1)$ to the standard 
N=2 superspace. The $SU(2)_R$ automorphisms are realized in PSS by the 
(non-linear) projective transformations. Though the HSS approach is the most 
symmetric and universal one, it also implies the infinite number of auxiliary 
fields, e.g. in the off-shell formulation of the FS hypermultiplet. This makes
 the relation between the HSS superfields and the component approach to be 
highly non-trivial. The PSS approach can be formulated with a finite number 
of auxiliary fields for a restricted class of N=2 NLSM, by assuming a 
holomorphic (polynomial) dependence upon the $CP(1)$ coordinate.  Allowing 
a more general (e.g., meromorphic) dependence of projective N=2 multiplets
upon the $CP(1)$ coordinate makes the PSS method to be essentially 
equivalent to the HSS one \cite{kuzp} (see Appendix B for a technical 
inroduction into PSS, and Appendix C for a technical introduction into HSS). 

The PSS construction of the N=2 NLSM metrics can be summarized into a short
prescription known as the {\it generalized Legendre transform}
\cite{lroc}. One considers sections of $O(2p)$ line bundles over
$CP(1)$, defined by all holomorphic polynomials of the $CP(1)$ projective
coordinate $\x$, of order $2p$ ({\it cf.} eq.~(B.2)),
$$ Q_{(2p)}(\x) = z + v\x + w_2\x^2 + \ldots + w_{2p-2}\x^{2p-2} +
(-1)^{p-1}\bar{v}\x^{2p-1} + (-1)^p\bar{z}\x^{2p}~,\eqno(2.32)$$
subject to the reality condition ({\it cf.} eq.~(2.30))
$$\Bar{Q_{(2p)}(\x)} =(-1)^p\bar{\x}^{2p}Q_{(2p)}(-1/\bar{\x})~.
\eqno(2.33)$$
One introduces the contour integral ({\it cf.} eqs.~(B.10) and (B.25))
$$F = \fracmm{1}{2\p i} \oint_C d\x \,\x^{-2} G(Q(\x),\x) \eqno(2.34)$$
in terms of a holomorphic function $G(Q(\x),\x)$. A K\"ahler
potential $K$ of the hyper-K\"ahler metric associated with the 
holomorphic input $(G,C)$ is found by performing the complex Legendre 
transform with respect to $v$ and $\bar{v}$ \cite{lroc} ({\it cf.} eq.~(B.26)),
$$ K(z,\bar{z},u,\bar{u}) =
F(z,\bar{z},v,\bar{v},w_a)-uv-\bar{u}\bar{v}~,\eqno(2.35)$$
subject to
$$ u \equiv \fracmm{\pa F}{\pa v}~,\qquad  \bar{u} \equiv 
\fracmm{\pa F}{\pa\bar{v}}~,\eqno(2.36)$$
when simultaneously extremizing $F$ with respect to all $w_a$, where
$a=2,\ldots,2p-2$ ({\it cf.} eq.~(B.23),
$$\fracmm{\pa F}{\pa w_a}=0~.\eqno(2.37)$$ 

The generalized Legendre transform provides the very powerful method for an
explicit construction of hyper-K\"ahler metrics, especially after taking into
account all $O(2p)$ sections over $CP(1)$ with $p=1,2,\ldots,\infty$. In
fact, one has to take all of them in the most general case. However, because 
of the complicated (highly non-linear) algebraic relations associated with the 
generalized Legendre transform, it seems to be very difficult to classify all
hyper-K\"ahler metrics (e.g., according to their isometries) by using this 
method. 

Harmonic superspace (HSS) can be independently justified by `relaxing' 
(i.e. lifting off-shell) the Fayet-Sohnius hypermultiplet constraints 
(2.29) in the on-shell case of $n=1$. It can be achieved with the 
{\it infinite} chain, 
$$  D^{(i}_{\a}L^{j)}=D_{\a k}L^{ijk}~,\quad 
\bar{D}^{(i}_{\dt{\a}}L^{j)}=D_{\dt{\a} k}L^{ijk}~,$$
$$D^{(i}_{\a}L^{jkl)}=D_{\a m}L^{ijklm}~,\quad 
\bar{D}^{(i}_{\dt{\a}}L^{jkl)}=D_{\dt{\a} m}L^{ijklm}~,$$
$$D^{(i}_{\a}L^{jklm)}=D_{\a n}L^{ijklmn}~,\quad 
\bar{D}^{(i}_{\dt{\a}}L^{jklm)}=D_{\dt{\a} n}L^{ijklmn}~,$$
$$ \cdots \eqno(2.38)$$
which involves all complex projective multiplets $L^{(i_1\cdots i_{2p+1})}$ 
up to $p=\infty$. The hypermultiplet HSS superfield $q^+$ (see
Appendix C) is equivalent to eq.~(2.38). Though it is definitely
possible to incorporate all relaxed projective supermultiplets,
including the one of eq.~(2.38), into the framework of the PSS 
construction (Appendix B), it is hardly convenient in practice.  
 
\subsection{Isometries of N=2 NLSM}

As regards the general {\it bosonic} NLSM (2.4), its isometry
(symmetry) Lie group $G$ may have an isotropy subgroup $H$ consisting of those
transformations of $G$ that leave a point $\{A^i\}$ in the NLSM target
space fixed. The remaining symmetries of $G$ move the point $\{A^i\}$,
being {\it non-linearly} realised. The infinitesimal action of $G$ reads
$$ \d A^i= \l^Mk_M^i(A)~,\eqno(2.39)$$
where $k_M^i(A)$, $M=1,2,\ldots,{\rm dim}\,G$, are the Killing vectors
generating the group $G$ and satisfying the Killing equation
$k^{(i;j)}_M=0$, while $\l^M$ are (Lie algebra) parameters. 
The isotropy subgroup $H(A)$ depends upon the point $\{A^i\}$ 
chosen. In adapted local coordinates, associated
with a given point $A^i$, the group $H(A)$ acts linearly, i.e. 
$$  \d A^i =i \l^X(T_X)\ud{i}{j}A^j~,\eqno(2.40)$$
where $T_X$ are the hermitian generators of $H(A)$. We follow 
ref.~\cite{haklr} here.
 
A description of isometries of a K\"ahler manifold has some special 
features related to the invariance of its complex structure under 
holomorphic transformations that do not mix $A^i$ and $\bar{A}_j$.
It is, therefore, natural to distinguish between the {\it holomorphic} 
isometries possessing the same property, and the isometries that do not. 
The Lie derivative associated with a  holomorphic Killing vector leaves  
both the K\"ahler metric {\it and} the complex structure invariant \cite{bag}. 
The action of the holomorphic isotropy subgroup in adapted coordinates reads
$$ \d A^i=i\l\ud{i}{j}A^j~,\quad \d\bar{A}_i=-i\bar{A}_j\l\ud{j}{i}~,\quad
{\rm where}\quad \l\ud{i}{j}=\l^X(T_X)\ud{i}{j}~.\eqno(2.41)$$

The K\"ahler potential is generically invariant under the isometry 
modulo a K\"ahler gauge transformation, 
$$ \d K =\h(A) +\bar{\h}(\bar{A})~,\eqno(2.42)$$
with a holomorphic function $\h(A)$. In the isotropic case one can 
always choose the K\"ahler gauge where $\h$ vanishes, so that
$$ \d K= i\l\ud{i}{j}(K_iA^j -K^j\bar{A}_i)=0~.\eqno(2.43)$$

To describe general holomorphic isometries, one introduces Killing 
vectors $k_a$ with holomorphic components $k_a^i(A)$ and their complex
conjugates $\bar{k}_{ai}(\bar{A})$ \cite{haklr},
$$\d A^i= \l^ak_a^i \equiv L_{\l\cdot k}A^i~,\qquad
\d\bar{A}_i= \l^a\bar{k}_{ai} \equiv L_{\l\cdot k}\bar{A}_i~.
\eqno(2.44)$$
In adapted coordinates, eq.~(2.44) takes the form of eq.~(2.41). The 
holomorphic and antiholomorphic components of the Killing vectors generate
two separate isometry algebras, and, in general coordinates, obey the 
Killing equations  
$$ K\ud{i}{j}k^j_{a;k}+K\ud{j}{k}\bar{k}\du{aj}{;i}=0~.\eqno(2.45)$$
Eq.~(2.45) implies the existence of the Killing potential associated with
a holomorphic Killing vector \cite{bag,haklr}.

Since a hyper-K\"ahler metric is fully characterized by its
quaternionic or {\it triholomorphic} structure, it is also quite natural
to distinguish between the {\it triholomorphic} isometries preserving
all three complex structures, and the non-triholomorphic isometries 
that do not share this property. The triholomorphic isometries are
also known in the literature as {\it translational}, while they
commute with N=2 supersymmetry in the N=2 NLSM. The non-triholomorphic 
isometries are sometimes called {\it rotational} since the action of the 
corresponding Killing vector on the complex structures amounts to
their rotation, while they do not commute with N=2 supersymmetry. 

In real coordinates (Appendix A), a triholomorphic Killing vector $k^m$
satisfies the equation
$$ P\du{\pm i}{j}\left( P_{\mp m}^{n}k^m\right)_{;j}=0~,\eqno(2.46)$$
where we have used the projection operators of eq.~(A.2). Eq.~(2.46) is
equivalent to the vanishing Lie derivative of each complex structure $J$,
$$ L_k J\du{i}{j}\equiv k^mJ\dud{i}{j}{,m}-k\ud{j}{,m}
J\du{i}{m}+k\ud{m}{,i}J\du{m}{j}=0~.\eqno(2.47)$$
In the special coordinates, where the complex structures have the form 
(A.6) and (2.13), and the Killing vector $k^m$ are manifestly 
holomorphic with respect to the canonical complex structure, the 
triholomorphicity condition reads \cite{haklr}
$$ \bar{\O}^{ij}\bar{k}\du{j}{;m}-
\bar{\O}^{jm}\bar{k}\ud{i}{;j}=0~,\eqno(2.48)$$
or, equivalently,
$$ \bar{\o}^{j[i}\bar{k}\du{j}{;m]}=0~.\eqno(2.49)$$

The triholomorphic condition (2.46) or (2.47) can be
considered as the integrability condition for the existence 
of a real {\it Killing potential} $X^{(J)}$ for each complex 
structure $J$, which satisfies the differential equation \cite{haklr}
$$ k^iJ_{ij}=-X\ud{(J)}{,j}~~.\eqno(2.50)$$
In the special coordinates, the existence of a real Killing potential 
$X^{(J)}$ amounts to the existence of a {\it holomorphic} Killing
potential $P$ and an {\it antiholomorphic} Killing potential $\bar{P}$, which
are defined with respect to $J^{(1)}\mp iJ^{(2)}$, respectively~\cite{haklr}.

Triholomorphic isometries of a hyper-K\"ahler metric significantly simplify
an explicit construction of the metric. Given an N=2 tensor multiplet
amongst the arguments of the PSS construction (Appendix B) or, equivalently, 
a section of the line bundle $O(2)$ in the generalized Legendre transform,
it always implies a translational (or triholomorphic) isometry. This can be
understood as the result of dualization of the conserved vector amongst 
the components of the N=2 tensor multiplet. The translational isometry is 
manifest in the corresponding K\"ahler potential of eq.~(B.13). This property
is not shared by the higher projective multiplets associated with the $O(2j)$ 
bundles over $CP(1)$ with $j>1$.

In the HSS approach, an analytic hypermultiplet superfield  $q^+$ and 
its analytic conjugate $\sbar{q}{}^+$ naturally form an $Sp(1)$
doublet, $q^{a+}$, where $a=1,2$. The free hypermultiplet action (C.7) 
has the manifest $Sp(1)$ isometry, in addition to the manifest 
$SU(2)_R$ isometry.  The $U(1)$ subgroup of the $Sp(1)$ rotations is given by 
$$ q^+ \to e^{i\a}q^+\quad {\rm and}\quad  
\sbar{q}{}^+ \to e^{-i\a}\sbar{q}{}^+~,\eqno(2.51)$$ 
and it can be identified with the $U(1)_R$ symmetry.

In order to get the most general N=2 NLSM in HSS, let's make the same trick
as in general relativity, where one goes from a flat space metric 
in general coordinates to the truly curved space metric. In HSS we can
assign extra $Sp(n)$ indices to the HSS superfields,
$$ q^{a+} ~\to~ q^{A+}~,\quad A=1,2,\ldots 2n~,\eqno(2.52)$$
and apply a reparametrization,
$$ q^{A+} \to {q^{A+}}' = f^{A+}(q,u)~,\quad {\rm with} \quad
u^{\pm i}\quad {\rm inert}~,\eqno(2.53)$$
to the free HSS action (Appendix C)
$$S_{\rm free}[q^A] =\fracmm{1}{2}
 \int_{\rm analytic} q_A^+ D^{++}q^{A+}~.\eqno(2.54)$$
One easily finds that the free action (2.54) gets transformed into 
$$ S[q^A]= \fracmm{1}{2}\int_{\rm analytic}\left\{ 
F^+_A(q,u)D^{++}q^{A+} + G^{(+4)}(q,u)\right\}\equiv 
\fracmm{1}{2}\int_{\rm analytic}\Lag^{(+4)}(q,u)~,\eqno(2.55)$$
with the {\it particular} functions $F^+_A$ and $G^{(+4)}$ given by
$$ F^+_A= f^+_B\fracmm{\pa f^{B+}}{\pa q^{A+}} \quad{\rm and}\quad
G^{(+4)}=f^+_B\pa^{++}f^{B+}~.\eqno(2.56)$$
It is therefore eq.~(2.56), with {\it arbitrary} complex functions
$F^+_A(q,u)$ and a {\it generic} real function $G^{(+4)}(q,u)$, that represents
the most general N=2 NLSM in HSS ({\it cf.} ref.~\cite{beg}). The action 
(2.56) is invariant under infinitesimal field reparametrizations, 
$$ \d q^{A+}=\r^{A+}(q,u)~,\quad \d u^{\pm}_i=0~,\eqno(2.57)$$ 
provided that 
$$ \d F^{+}_A= F^+_B\fracmm{\pa\r^{B+}}{\pa
q^{A+}} \quad{\rm and}\quad  \d G^{(+4)}= F^+_A\pa^{++}\r^{A+}~.\eqno(2.58)$$
The HSS `vielbein' $F_A$ is the pure gauge field with respect
to the HSS reparametrizations, and it can be gauge-fixed to the 
`canonical' form, $F^+_A=q^+_A$, in adapted coordinates on the NLSM
target space. It results in the standard form of the most general N=2 NLSM
action in HSS \cite{gios,gios2},
$$ S_{\rm NLSM}[q]= \fracmm{1}{2}\int_{\rm analytic}\left\{
q^+_AD^{++}q^{A+} + \ck^{(+4)}(q,u)\right\}~.\eqno(2.59)$$
The function $\ck^{(+4)}(q,u)$ is called a {\it hyper-K\"ahler potential}
 \cite{gios,gios2}. 

Since isometries are the symmetries of the NLSM action, not of the NLSM
Lagrangian, in HSS the latter may vary into a total harmonic derivative,
$\d \Lag^{(+4)}=D^{++}\L^{++}(q,u)$, because of the identity 
$$\int_{\rm analytic} D^{++}\L^{++}\equiv \int_{\rm analytic}\left[
\pa^{++}\L^{++} +\fracmm{\pa\L^{++}}{\pa q^{A+}}D^{++}q^{A+}\right]=0~.
\eqno(2.60)$$
The action (2.55) is invariant under an infinitesimal isometry transformation
$$ \d q^{A+}=\ve^X \r^{XA+} \eqno(2.61)$$
with some constant parameters  $\ve^X$, $X=1,2,\ldots,{\rm dim}\,H$, and
the triholomorphic Killing vectors $\r^{XA+}(q,u)$  provided that \cite{beg} 
$$\eqalign{
 \left( \fracmm{\pa F^+_A}{\pa q^{B+}} +
\fracmm{\pa F^+_B}{\pa q^{A+}}\right)\r^{XB+}=~& \fracmm{\pa\L^{X++}}{\pa
 q^{A+}}~,\cr
\left( \fracmm{\pa V^{(+4)}}{\pa q^{A+}}-\pa^{++}F^+_A\right)
\r^{XA+}=~& \pa^{++}\L^{X++}~.\cr}\eqno(2.62)$$
In adapted coordinates eq.~(2.62) simplifies to
$$  \fracmm{\pa\L^{X++}}{\pa q^{A+}}=-2\r^{X+}_A~,\quad
-2\pa^{++}\L^{X++}= \fracmm{\pa\ck^{(+4)}}{\pa q^+_A}
\fracmm{\pa\L^{X++}}{\pa q^{A+}}~.\eqno(2.63)$$
By analogy with the N=1 superspace description of N=2 NLSM isometries, 
the analytic HSS superfield $\L^{X++}$ is called the (triholomorphic) Killing
potential of the (triholomorphic) Killing vector $\r^{XA+}$ in HSS \cite{beg}.

If the N=2 NLSM action (2.59) in adapted coordinates has a linearly realised
isometry, eq.~(2.61) takes the form ({\it cf.} eqs.~(2.40) and (2.41))
$$ \d q^{A+}= i\ve^X(T_X)\ud{A}{B}q^{B+}~,\eqno(2.64)$$
where $T_X$ are the generators of $H$. The corresponding Killing vectors are
linear in $q$, whereas their Killing potentials are quadratic \cite{beg},
$$ \L^{X++}=-iq^+_A(T_X)\ud{A}{B}q^{B+}~.\eqno(2.65)$$

\section{Exact dynamics of a single hypermultiplet}

The general 4d, N=2 NLSM Lagrangian for a single hypermultiplet in HSS
reads (subsect.~2.3)
$$ -\cl^{(+4)}=~ \sbar{q}{}^+D^{++}q^+ +
\ck^{(+4)}(\sbar{q}{}^{+},q^+;u^{\pm})~,\eqno(3.1)$$
where the hyper-K\"ahler potential $\ck^{(+4)}$ is an arbitrary
function of an unconstrained analytic HSS superfield $q^+$, its 
analytic conjugate $\sbar{q}{}^+$, and harmonics $u_i^{\pm}$. A function
$\ck^{(+4)}$ should be of $U(1)$ charge $(+4)$ in order to cancel the 
opposite $U(1)$ charge of the analytic measure in HSS (Appendix C). 
As regards the hypermultiplet LEEA also having the form of eq.~(3.1), its
hyper-K\~ahler potential plays the role similar to that of the 
holomorphic Seiberg-Witten potential $\cf$ in the abelian N=2 vector
LEEA \cite{sw}. Because of manifest N=2 supersymmetry of eq.~(3.1) 
describing the propagating Fayet-Sohnius hypermultiplet degrees of freedom 
only, the equations of motion for the HSS action (3.1) determine (at least,
 in principle) the component hyper-K\"ahler NLSM metric in terms of a 
single HSS potential $\ck^{(+4)}$. It is not known how to deduce an 
explicit NLSM metric from eq.~(3.1) in the case of a generic 
hyper--K\~ahler potential, though some explicit examples are available
(Appendix C).

A crucial simplification arises when the $SU(2)_R$ symmetry is not
broken, which is expected to be the case for the hypermultiplet LEEA 
(sect.~1). Since the $SU(2)_R$ transformations are linearly realised 
in HSS, the $SU(2)_R$ isometry of the hypermultiplet LEEA just means 
that the corresponding hyper-K\"ahler potential $\ck^{(+4)}$ should be 
independent upon harmonics. This observation gives rise to the most 
general invariant `Ansatz' for the LEEA of a single hypermultiplet, 
in the form of the most general, of $U(1)$ charge $(+4)$, harmonic-indepenent
hyper-K\"ahler potential ({\it cf.} ref.~\cite{gios}),
$$ \ck^{(+4)}= \fracm{\l}{2}(\sbar{q}{}^+)^2(q^+)^2+\left[ 
\g \sbar{(q^+)}{}^4 +  \b\sbar{(q^+)}{}^3 q^+ +{\rm h.c.}\right]~,\eqno(3.2)$$
with one real $(\l)$ and two complex $(\b,\g)$ parameters. The 
$Sp(1)=SU(2)_{\rm PG}$ transformations of $q^+_a$ leave the form of eq.~(3.2) 
invariant but not the coefficients. Since $SU(2)_{\rm PG}$ is the symmetry
of a free hypermultiplet action (C.7), it can be used to reduce the
number of coupling constants in the family of hyper-K\"ahler metrics 
associated with the hyper-K\"ahler potential (3.2) from five to two. In
addition, eq.~(3.2) implies the conservation law \cite{gios} 
$$ D^{++}\ck^{(+4)}=0\eqno(3.3)$$
that is valid on the equations of motion of the hypermultiplet HSS superfield,
$$D^{++}\sbar{q}{}^+ =\pa\ck^{(+4)}/\pa q^+\quad {\rm and}\quad 
D^{++}q^+ =-\pa\ck^{(+4)}/\pa\sbar{q}{}^+~.\eqno(3.4)$$

\subsection{Hypermultiplet LEEA in the Coulomb branch}

The manifestly N=2 supersymmetric HSS description of the hypermultiplet 
LEEA allows us to exploit the constraints imposed by unbroken N=2 
supersymmetry and its automorphism  symmetry in the very efficient 
and transparent way. For example, as regards a perturbation 
theory in 4d, N=2 supersymmetric QED (or in the Coulomb branch of N=2
supersymmetric $SU(2)$ Yang-Mills theory \cite{sw}), the unbroken
symmetry is given by 
$$ SU(2)_{R,~{\rm global}}\times U(1)_{\rm local}~.\eqno(3.5)$$

The {\it unique} hypermultiplet self-interaction consistent with N=2 
supersymmetry and the symmetry (3.5) in HSS is described by the 
hyper-K\"ahler potential 
$$ \ck^{(+4)}_{\rm Taub-NUT}=\fracmm{\l}{2}\left(\sbar{q}{}^+q^+\right)^2~,
\eqno(3.6)$$
just because this is the only function of $U(1)$ charge $(+4)$ that is 
independent upon harmonics, being invariant under the $U(1)$ phase
transformations (2.51) too.  

\begin{figure}
\vglue.1in
\makebox{
\epsfxsize=4in
\epsfbox{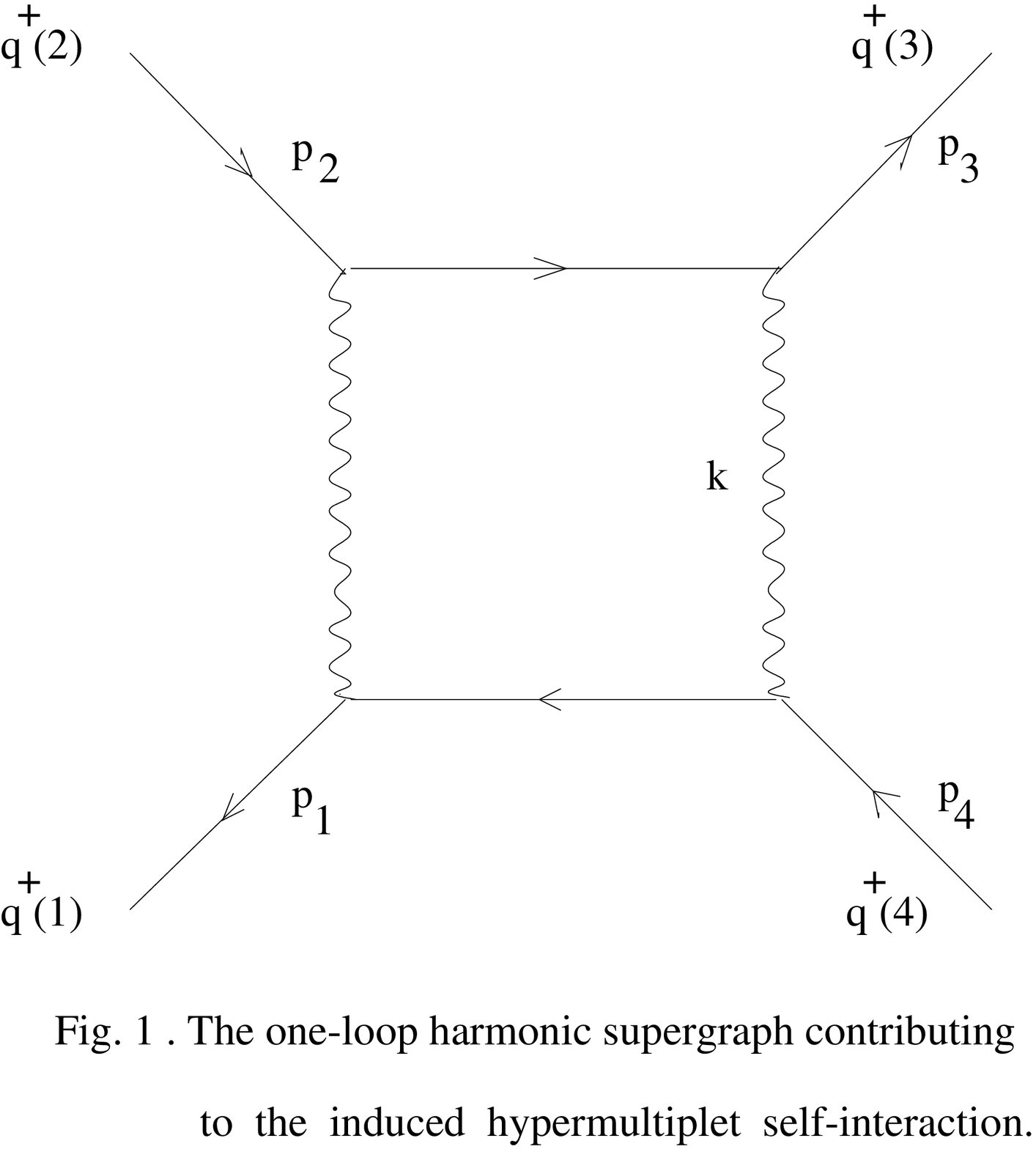}
}
\end{figure}

The induced coupling constant $\l$ of eq.~(3.6) in the one-loop approximation
can be determined from a calculation of the HSS graph shown in Fig.~1, {\it 
after taking into account central charges} \cite{ikz}. The analytic propagator
 (the wave lines in Fig.~1) of the N=2 vector HSS superfield $V^{++}$ in N=2 
supersymmetric Feynman gauge is given by \cite{hssf} 
$$ i\VEV{ V^{++}(1)V^{++}(2)}=\fracmm{1}{\Box_1}(D_1^+)^4\d^{12}(\cz_1-\cz_2)
\d^{(-2,2)}(u_1,u_2)~,\eqno(3.7)$$
where the harmonic delta-function $\d^{(-2,2)}(u_1,u_2)$ has been 
introduced \cite{hssf}. The hypermultiplet analytic propagator (the
solid lines in Fig.~1) with {\it non-vanishing} central charges is 
more complicated \cite{ikz},~\footnote{Eq.~(3.8) reduces to the HSS 
hypermultiplet propagator found in ref.~\cite{hssf} when $Z=0$.}
$$ i\VEV{q^+(1)q^+(2)}=\fracmm{-1}{\Box_1+m^2}
\fracmm{(D^+_1)^4(D^+_2)^4}{(u^+_1u_2^+)^3}e^{\t_3[v(2)-v(1)]}\d^{12}
(\cz_1-\cz_2)~,\eqno(3.8)$$
where we have used the `pseudo-real' $Sp(1)$ notation, see eq.~(C.7). The
`bridge' $v$ satisfies an equation ${\cal D}^{++}e^v=0$, whereas
 $m^2=\abs{Z}^2$ is the hypermultiplet BPS mass. One finds 
$$ iv =-Z(\bar{\theta}^+\bar{\theta}^-)-\bar{Z}(\theta^+\theta^-)~.\eqno(3.9)$$
A calculation of the LEEA from the one-loop HSS graph in Fig.~1 is 
straightforward, while $Z\neq 0$ is essential \cite{ikz}. We find the 
predicted form (3.6) of the induced hyper-K\"ahler potential with the one-loop
 induced NLSM coupling constant 
$$ \l=\fracmm{g^4}{\p^2}\left[ \fracmm{1}{m^2}\ln\left( 1+\fracmm{m^2}{\L^2}
\right) -\fracmm{1}{\L^2+m^2}\right] \eqno(3.10)$$
in terms of the abelian coupling constant $g$, the bare BPS mass $m^2$ and 
the IR-cutoff $\L$. Note that $\l\neq 0$ only when $Z\neq 0$. The naive 
`non-renormalization theorem' forbids the appearance of the quantum 
corrections given by integrals over a subspace of the full N=2 superspace,
like the one in eq.~(3.6). However, this `non-renormalization theorem'
does not apply in the case under consideration, because of the 
non-vanishing central charge $Z$ ({\it cf.} ref.~\cite{hssf}).

The infra-red divergence of the one-loop induced effective coupling $\l$ 
in eq.~(3.10) may cause concern about the consistency of our approach.
It is worth mentioning, however, that eq.~(3.6) is manifestly N=2 
supersymmetric at {\it any~} $\l$. In other words, the infra-red
cutoff is consistent with N=2 supersymmetry of the hypermultiplet
LEEA. We expect the IR divergences to disappear after summing up 
(IR-divergent) higher-loop Feynman supergraphs with four external FS 
hypermultiplets  (Fig.~2).  

\begin{figure}
\vglue.1in
\makebox{
\epsfxsize=4in
\epsfbox{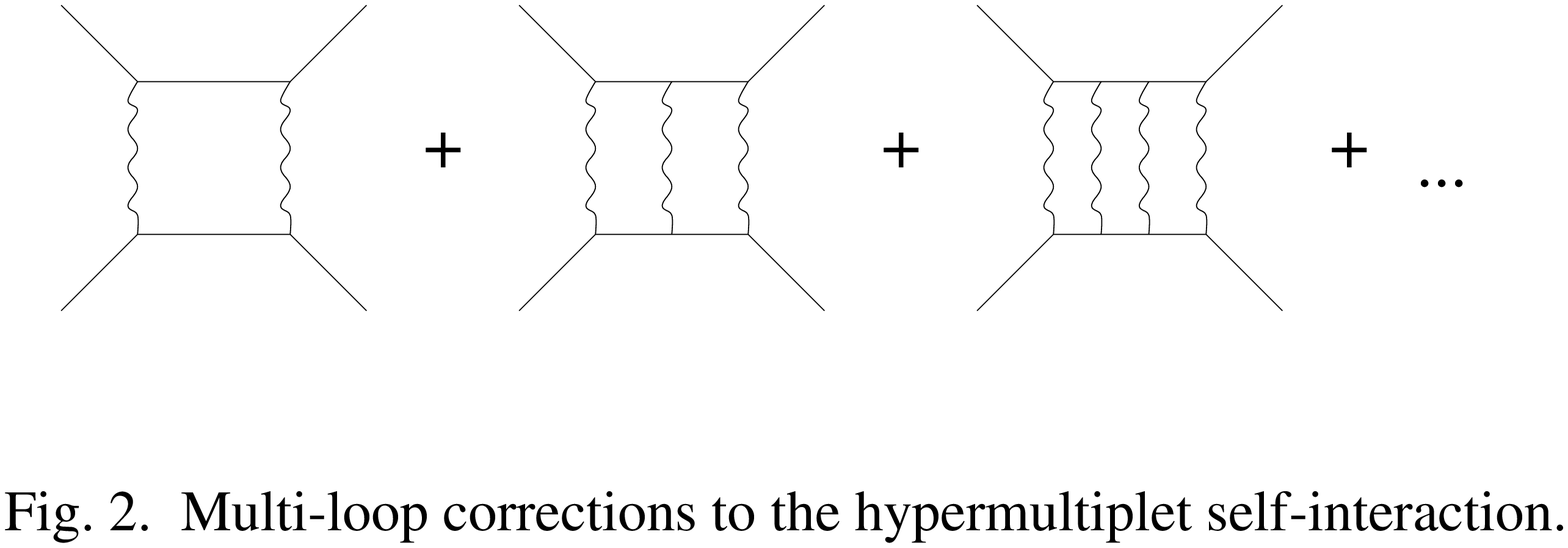}
}
\end{figure}

It seems to be rather difficult to calculate the exact dependence 
of $\l$ upon the fundamental parameters of the underlying gauge
theory, such as the Yang-Mills coupling constant, an N=2 central
charge and a Higgs expectation value, by using quantum perturbation
theory. The perturbative expansion in terms of Feynman graphs assumes 
the validity of the weak-coupling description associated with a
particular choice of fields and coupling constants in (a portion of) 
the quantum moduli space, whereas the parameter $\l$ is essentially
non-perturbative ({\it cf.} the Fermi constant $F_{\p}$ in QCD).
Nevertheless, the exact geometrical (Taub-NUT) nature of the result
(3.6), associated with a fundamental monopole belonging to the 
hypermultiplet under consideration, is very clear in the HSS
approach. The coefficient $\l$ is simply related to the Taub-NUT
mass, $\l=\frac{1}{4}M^{-2}$ (see Appendix C).  

To understand the hyper-K\"ahler geometry associated with the 
hyper-K\"ahler potential (3.6), one may perform an N=2 supersymmetric 
reduction of the FS hypermultiplet to an N=2 {\it tensor} multiplet, 
and then rewrite the corresponding dual HSS action into PSS. Unlike the 
off-shell FS hypermultiplet, the off-shell N=2 tensor multiplet has
the finite number of the auxiliary fields --- see eq.~(2.31). 
The N=2 tensor multipet constraints (2.28) can be rewritten to HSS as 
$$ D^{++}L^{++}=0 \quad {\rm and} \quad \sbar{L}{}^{++}=L^{++}~,\eqno(3.11)$$
where $L^{++}=u^+_iu^+_jL^{ij}(\cz)$. Let's substitute  
 (we temporarily set $\l=1$)
$$ \ck^{(+4)}_{\rm TN}=\fracm{1}{2}(\sbar{q}{}^+q^+)^2
=-2(L^{++})^2~,\quad {\rm or,~equivalently,}\quad 
\sbar{q}{}^+q^+=2iL^{++}~,\eqno(3.12)$$
which is certainly allowed because of eq.~(3.3). The constraints (2.28) can be 
taken into account off-shell, by using an extra real analytic HSS
superfield $\o$ as the Lagrange multiplier. Changing the variables from 
$(\sbar{q}{}^+,q^+)$ to $(L^{++},\o)$ amounts to the N=2 duality 
transformation in HSS. An explicit solution to eq.~(3.12) is known \cite{gio},
$$ q^+=-i\left(2u^+_1+if^{++}u^-_1\right)e^{- i\o/2}~,
\quad \sbar{q}{}^+=i\left( 2u^+_2-if^{++}u^-_2\right)e^{i\o/2}~,\eqno(3.13)$$
where the function $f^{++}$ is given by
$$ f^{++}(L,u) = \fracmm{2(L^{++}-2iu_1^+u_2^+)}{1+\sqrt{1-4u^+_1u^+_2
u^-_1u^-_2 -2iL^{++}u_1^-u_2^-}}~~~.\eqno(3.14)$$

It is straightforward to rewrite the free (massless) HSS action
(C.7) in terms of the new variables. This results in the {\it improved} 
(i.e. N=2 superconformally invariant) N=2 tensor multiplet action \cite{gio},
$$S\low{\rm improved} = \ha \int d\z^{(-4)}du (f^{++})^2~.\eqno(3.15)$$
The action dual to the NLSM action defined by eq.~(C.10) is,
therefore, given by a sum of the non-improved (quadratic) and improved 
(non-polynomial) HSS actions for the N=2 tensor multiplet \cite{hklr,gio},
$$ S_{\rm Taub-NUT}[L;\o]= S\low{\rm improved} 
+\ha \int d\z^{(-4)}du \left[(L^{++})^2 +\o D^{++}L^{++}\right]~.\eqno(3.16)$$

To understand the peculiar structure of the improved action defined 
by eqs.~(3.14) and (3.15), let's extract a constant `vacuum expectation value'
 $c^{ij}$ out of $L^{ij}$ by rewriting it to the form  
$$ L^{++}(\z,u)=c^{++} + l^{++}(\z,u)~.\eqno(3.17)$$
We use the notation 
$$c^{\pm\pm}=c^{ij}u_i^{\pm}u_j^{\pm}~,\quad 
\Bar{(c^{ij})}=\ve_{ik}\ve_{jl}c^{kl}~,\quad c^2=\ha c_{ij}c^{ij}\neq 0~,$$
$$ f^{++}(L,u)\equiv l^{++}f(y)~,\quad y=l^{++}c^{--}~.\eqno(3.18)$$
The function $f(y)$ then appears to be a solution to the quadratic equation,
$$ \fracmm{1}{f(y)}= 1 +\fracmm{yf(y)}{4c^2}~.\eqno(3.19)$$
It can be shown that this equation follows from the rigid N=2 superconformal 
invariance on the improved action \cite{gio}. The improved {\it action},
 defined by eqs.~(3.15) and (3.19), does not really depend upon $c^{ij}$ 
because of its $SU(2)_{\rm conf.}$ invariance. 

It is also straightforward to rewrite eq.~(3.16) to N=2 PSS and even 
to N=1 superspace, where it takes the form of eqs.~(B.3) and (B.10), 
respectively. After restoring the dependence upon $\l\equiv\frac{1}{4}M^{-2}$
in eq.~(3.16), we thus reproduce the PSS action of ref.~\cite{klr}, with 
$$ \oint G = M \oint_{C_0} \fracmm{Q_{(2)}^2}{2\x} + 
\oint_{C_r} Q_{(2)}(\ln Q_{(2)}-1)~,\eqno(3.20)$$
where the contour $C_0$ goes around the origin, whereas the contour
$C_r$ encircles the roots of a quadratic equation, 
$$ Q_{(2)}(\x)=0~, \eqno(3.21)$$ 
in complex $\x$-plane. The hyper-K\"ahler metric of the N=2 NLSM 
defined by eqs.~(3.16) and (3.20) is equivalent to the Taub-NUT metric 
after the N=1 superspace Legendre transform (see Appendix B)
\cite{hklr}, with the mass parameter $M=\ha\l^{-1/2}$. Stated
differently, eq.~(3.6) describes the hyper-K\"ahler potential of
the Taub-NUT metric in HSS (Apendix C).

The known $U(2)=SU(2)\times U(1)$ isometry of the Taub-NUT metric is 
clearly consistent with eq.~(3.5). It is instructive to investigate 
the realization of this internal symmetry in the various formulations 
of the N=2 Taub-NUT NLSM mentioned in this paper. The hyper-K\"ahler potential
(3.6) in terms of the FS hypermultiplet $q^+$ provides 
the manifestly invariant formulation. In the dual HSS form, 
in terms of $(L^{++},\o)$, the non-abelian factor $SU(2)$ is
represented by the $SU(2)_{\rm conf.}$, whereas the abelian factor $U(1)$ is 
realized by constant shifts of $\o$. The $SU(2)_R$ transformations act 
in PSS in the form of {\it projective} (fractional) transformations (B.7).

Eqs.~(B.8) and (B.9) imply that the $SU(2)_R$ invariant PSS potential 
$G(Q_{(2)})$ should be `almost' linear in $Q_{(2)}$, 
like in the second term of eq.~(3.20) where the extra logarithmic
factor is merely responsible for defining the (closed) integration contour. 
The transition $u_i\to\x_i=(1,\x)$ describes a holomorphic projection of 
HSS to PSS where the analytic superfield $L^{++}(\z,u)$ is replaced by a
holomorphic (with respect to $\x$) section $Q_{(2)}(L,\x)$ of the line 
bundle $O(2)$ whose fiber is parametrized by the constrained N=1
superfields, $\c$ and $g$. The equation $y^2=\c -i\x g +\x^2\bar{\c}$ 
defines the Riemann sphere in ${\bf C}^2$ parametrized by 
$(y,\x)$, where $y^2\equiv\left.Q_{(2)}\right|$.

\subsection{Hypermultiplet LEEA in the Higgs branch}

There are no instantons in an abelian N=2 supersymmetric quantum field theory.
This means that the perturbative result of subsect.~3.1 about the 
hypermultiplet LEEA described by the Taub-NUT metric is, in fact, {\it exact}
in the abelian case. If, however, the underlying N=2 gauge field theory has 
a non-abelian gauge group whose rank is larger than one (say, $SU(3)$), one 
may expect nonperturbative contributions to the hypermultiplet LEEA 
(in the Higgs branch) from instantons and anti-instantons, which break the 
$U(1)$ symmetry in eq.~(3.5) ({\it cf.} ref.~\cite{hwit}). 

Given the most general $SU(2)_R$-invariant hyper-K\"ahler 
potential (3.2), let's make a substitution \cite{prep}
$$ \ck^{(+4)}(q,\sbar{q})\equiv\fracm{\l}{2}(\sbar{q}{}^+)^2(q^+)^2+\left[ 
\g\sbar{(q^+)}{}^4 + \b\sbar{(q^+)}{}^3q^+ +{\rm h.c.}\right]
=L^{++++}(\z,u)~,\eqno(3.22)$$
where the real analytic superfield $L^{++++}$ satisfies the
conservation law (3.3), i.e.  
$$ D^{++}L^{++++}=0~.\eqno(3.23)$$
Eq.~(3.23) can be recognized as the {\it off-shell} N=2 superspace constraints
$$ D\low{\a}{}^{(i}L^{jklm)}=\bar{D}_{\dt{\a}}{}^{(i}L^{jklm)}=0~,\eqno(3.24)$$
where $L^{++++}=u^+_iu^+_ju^+_ku^+_lL^{ijkl}(\cz)$,  
while eq.~(3.22) implies the reality condition
$$ \Bar{L^{ijkl}}=\ve_{im}\ve_{jn}\ve_{kp}\ve_{lq}L^{mnpq}~,\eqno(3.25)$$
defining together the $O(4)$ projective N=2 supermultiplet (see subsect.~2.2.
and Appendix B). Unlike the $O(2)$ tensor multiplet, the $O(4)$ multiplet 
does not have a conserved vector (or a gauge antisymmetric tensor) 
amongst its field components, which implies the absence of the $U(1)$ 
triholomorphic isometry in the N=2 NLSM to be constructed in terms 
of $L^{++++}$. 

The N=2 supersymmetric PSS construction of the invariant actions
(Appendix B), in terms of a PSS potential $G(Q(\x),\x)$, equally 
applies to the projective $O(4)$ supermultiplets, while $L^{++++}$
should enter the universal PSS action (B.3) via the argument 
$$ Q_{(4)}(\cz,\x)=\x_{i}\x_{j}\x_{k}\x_lL^{ijkl}(\cz)~,\quad \x_i=(1,\x)~.
\eqno(3.26)$$
The N=1 superspace projections of eqs.~(3.26) and (B.3) are given by
eqs.~(B.24) and (B.25), respectively, in terms of the N=1 chiral
superfield $\c$, the N=1 complex linear superfield $W$, and the N=1
general (unconstrained) real superfield $V$.

The auxiliary field component $C$ of the $O(4)$ projective multiplet, 
defined by eq.~(2.31) (or, equivalently, the N=1 superfield $V$ in the
N=1 superspace reformulation) enters the PSS action (B.3) as
the Lagrange multiplier, whose elimination via its `equation of
motion' gives rise to an algebraic constraint, 
$$ {\rm Re}\,\oint \fracmm{\pa G}{\pa Q_{(4)}}=0~.\eqno(3.27)$$
Eq.~(3.27) reduces the number of independent physical real scalars
from five to four, which is consistent with the fact that the real 
dimension of any hyper-K\"ahler manifold is a multiple of four
(Appendix A). After solving the constraint (3.27), the complex 
linear N=1 superfield $W$ can be traded for yet another N=1 chiral 
superfield $\j$, by the use of the N=1 superfield Legendre transform. 
It results in the N=1 superspace K\"ahler potential 
$K(\c,\bar{\c},\j,\bar{\j})$ of the N=2 supersymmetric NLSM (Appendix B).

The most straightforward procedure of calculating the dependence
$q(L)$ out of the definition (3.22), as well as performing an explicit 
N=2 supersymmetric duality transformation of the free FS action (C.7) 
into the dual (improved) action of the constrained N=2 superfield $L^{ijkl}$
defined by eqs.~(3.24) and (3.25), needs the explicit roots of the
quartic polynomial. Though it is possible to calculate the roots, 
the results are not very illuminating. In fact, the explicit roots
are not even necessary to determine the explicit form of the dual N=2
PSS action. Within the manifestly N=2 supersymmetric approach, it is
the $SU(2)_R$ symmetry and regularity requirements that are sufficient 
to fix the action in question, either in HSS or PSS 
({\it cf.} ref.~\cite{gio}). The one real and two complex constants, 
$(\l,\b,\g)$, respectively, parametrizing the hyper-K\"ahler potential
(3.22) can be naturally united into an $SU(2)$ 5-plet $c^{ijkl}$
subject to the reality condition (3.25). After extracting a constant
piece out of $q^+$, say, $q^+_a=u^+_a + \tilde{q}^+_a$ and $u_a=(1,\x)$, 
and collecting all constant pieces on the left-hand-side of
eq.~(3.22), we can identify their sum with a constant piece
$c^{++++}=c^{ijkl}u^+_iu^+_ju^+_ku^+_l$ of $L^{++++}$ on the right-hand-side
of eq.~(3.22), representing the constant vacuum expectation values of
the N=1 superfield components of $L^{++++}$,
$$ \l=\VEV{V}~,\quad \b=\VEV{W}~,\quad\g=\VEV{\c}~.\eqno(3.28)$$

The $SU(2)_R$ transformations in PSS are the projective transformations 
(B.7), so that the PSS potential $G$ of the `improved' $O(4)$ projective
multiplet action having the form (B.3) must be proportional to
$\sqrt{Q_{(4)}}$ just because of the transformation rules  
$$G(Q_{(4)}'(\x'),\x')=\fracmm{1}{(a+b\x)^2}G(Q_{(4)}(\x),\x) 
\quad{\rm and}\quad Q_{(4)}'(\x')
=\fracmm{1}{(a+b\x)^4}Q_{(4)}(\x)~.\eqno(3.29)$$
The most general non-trivial contour $C_r$ in complex $\x$-plane, 
whose definition is compatible with the projective $SU(2)$ symmetry, 
is the one encircling the roots of the quartic ({\it cf.} eq.~(3.21)),
$$\left.Q_{(4)}(\x)\right|=p + \x q + \x^2 r -\x^3\bar{q}+\x^4\bar{p}~,
\eqno(3.30)$$ 
with one real $(r)$ and two complex $(p,q)$ additional parameters
belonging to yet another 5-plet of $SU(2)_{\rm PG}=Sp(1)$. The
projective $SU(2)$ invariance can be used to reduce the number of independent 
parameters in the corresponding family of the hyper-K\"ahler metrics 
from five to two, which is consistent with the HSS predictions.~\footnote{The
generalization of eq.~(3.21) similarly to eq.~(3.30) is `empty' since a 
quadratic polynomial \newline ${~~~~~}$ 
$c_{(2)}(\x)=p-i\x q+\x^2\bar{p}$ can always be removed by an $SU(2)$ 
transformation.} We didn't attempt to establish an explicit relation 
between the HSS coefficients $(\l,\g,\b)$ and the PSS coefficients 
$(r,q,p)$. The most natural contour $C_r$, surrounding roots of the equation
$$ Q_{(4)}(\x)=0~, \eqno(3.31)$$
just leads to the only non-singular hyper-K\"ahler NLSM metric 
(see subsect.~3.3).

Taking together the above considerations implies that the $SU(2)$-invariant 
PSS action, dual to the HSS action defined by eqs.~(3.1) and (3.2), 
is given by  
$$ \fracmm{1}{2\p i}\oint G =-\fracmm{1}{2\p i}\oint_{C_0} 
\fracmm{Q_{(4)}}{\x} + \oint_{C_r} \sqrt{Q_{(4)}}~.\eqno(3.32)$$
The constraint (3.27) reads in the case as follows:
$$ \oint_{C_r} \fracmm{d\x}{\sqrt{Q_{(4)}}}=1~.\eqno(3.33)$$
The generalized Legendre transform of the function (3.32) is known
\cite{iro}, so we can simply `borrow' some of the results of ref.~\cite{iro}
here.

Due to the reality condition (3.25), the quartic equation (3.31) has two
pairs of roots $(\r,-1/\bar{\r})$ related by an $SL(2,{\bf Z})$ 
transformation and satisfying the defining relation
$$ Q_{(4)}(\x)=
c(\x-\r_1)(\bar{\r}_1\x+1)(\x-\r_2)(\bar{\r}_2\x+1)~.\eqno(3.34)$$
The branch cuts in complex $\x$-plane can be chosen to run from
$\r_1$ to $-1/\bar{\r}_2$ and from $\r_2$ to $-1/\bar{\r}_1$. The
contour integration in eq.~(3.33) can then be reduced to the complete
elliptic integral (in the Legendre normal form) over the branch cut \cite{iro},
$$ \fracmm{4}{\sqrt{c(1+\abs{\r_1}^2)(1+\abs{\r_2}^2)}}
\int^1_0\fracmm{d\x}{\sqrt{(1-\x^2)(1-k^2\x^2)}}=1~,\eqno(3.35)$$
with the modulus
$$ k^2= \fracmm{(1+\r_1\bar{\r}_2)(1+\r_2\bar{\r}_1)}{(1+\abs{\r_1}^2)
(1+\abs{\r_2}^2)}~.\eqno(3.36)$$
The constraint (3.35) can be explicitly solved in terms of the
complete elliptic integrals, 
$$ K(k)=\int^{\p/2}_0 \,\fracmm{d\g}{\sqrt{1-k^2\sin^2\g}}~~,\quad
E(k) =\int^{\p/2}_0 d\g\,\sqrt{1-k^2\sin^2\g}~~,\eqno(3.37)$$
of the first and second kind, respectively, by using the following
parametrization \cite{iro}:
$$ \eqalign{ 
\F ~=~&  2e^{2i\vf} \left[ \cos(2\j) (1+\cos^2\vq)\right. \cr 
& \left.+ 2i\sin(2\j)\cos\vq + (2k^2-1)\sin^2\vq \right] K^2(k)~,\cr
H ~=~&  8e^{i\vf}\sin\vq \left[ \sin(2\j) \right. \cr
& \left.- i\cos(2\j)\cos\vq + i(2k^2-1)\cos\vq \right] K^2(k)~,\cr
V ~=~&  4\left[ -3\cos(2\j)\sin^2\vq +(2k^2-1)(1-3\cos^2\vq)\right]
K^2(k)~,\cr}\eqno(3.38)$$
where the Euler `angles' $(\vq,\j,\vf)$ have been introduced. Together with 
the modulus $k$ they represent the independent (superfield) coordinates in 
the N=2 NLSM under consideration. 

Being applied to the function (3.32), the Legendre transform (2.35) with 
respect to $W$, on the constraint (2.37) having of the form (3.35), gives 
rise to the (double cover of) {\it Atiyah-Hitchin} (AH) space $M_2$ as the 
NLSM target space. It was demonstrated in ref.~\cite{iro} by comparing the 
hyper-K\"ahler structures, which is enough to claim the equivalence between 
the N=2 NLSM metrics in accordance with the general discussion in 
subsect.~2.1. 

The AH space $M_2$ was originally introduced as the (centered) moduli space of
two (fundamental) BPS $SU(2)$ monopoles \cite{ati}. The metric of the AH space
is known to be the only {\it regular} hyper-K\"ahler metric with the entirely 
non-triholomorphic $SO(3)$ symmetry rotating hyper-K\"ahler complex 
structures \cite{ati}. In the {\it Donaldson} description \cite{don} 
of the AH space (see subsection 4.2 for basic definitions), the AH
space $M_2$ is described by the quotient of an algebraic curve in ${\bf C}^3$,
$$ x^2 -zy^2=1~,\quad {\rm where}\quad x,y,z\in {\bf C}~,\eqno(3.39)$$
under $Z_2:~(x,y,z)\equiv (-x,-y,z)$. Eq.~(3.39) thus describes 
the $SU(2)$-symmetric universal (2-fold) covering $\tilde{M}_2$ 
of the AH space. Since our N=2 superspace techniques are purely local, 
we are unable to distinguish between $M_2$ and $\tilde{M}_2$. Accordingly,
we make no distinction between $SO(3)$ and $SU(2)$ in our 
considerations.~\footnote{See, however, ref.~\cite{sw3} for a discussion 
of the global issues associated with $M_2$ and $\tilde{M}_2$.}

The line element of any four-dimensional (Euclidean) metric having 
$SO(3)$ isometry can be written down in the Bianchi IX formalism as follows:
$$ ds^2 = f^2(t)dt^2 +A^2(t)\s^2_1 + B^2(t)\s^2_2 +C^2(t)\s^2_3~,
\eqno(3.40)$$
where $f(t)=\ha ABC$, while $\s_i$ stand for the $SO(3)$-invariant 1-forms
$$ \eqalign{
\s_1 ~=~ & +\frac{1}{2}\left(\sin\j d\vq - \sin\vq \cos\j d \vf\right)~,\cr
\s_2 ~=~ & -\frac{1}{2}\left(\cos\j d\vq + \sin\vq \sin\j d \vf\right)~,\cr
\s_3 ~=~ & +\frac{1}{2}\left(d\j + \cos\vq d\vf\right)~,\cr}\eqno(3.41)$$
in terms of the Euler angles $(\vq,\j,\vf)$. The one-forms (3.41) obey 
the relations
$$ \s_i \wedge \s_j = \frac{1}{2}\ve_{ijk} d\s_k~,\quad i,j,k=1,2,3~.
\eqno(3.42)$$

The standard parametrization of the AH metric uses the complete
elliptic integrals (3.37) and the modulus $k$ related to the variable
$t$ of eq.~(3.40) via the relation
$$ t=-\fracmm{2K(k')}{\p K(k)}~,\eqno(3.43)$$
where $k'$ is known as the complementary modulus, ${k'}^2=1-k^2$. 
The AH metric in terms of
the independent coordinates $(k;\vq,\j,\vf)$ reads \cite{ati}
$$ ds^2_{\rm AH}=\fracmm{1}{4}A^2B^2C^2\left(\fracmm{dk}{k{k'}^2K^2}\right)^2
+ A^2(k)\s_1^2 + B^2(k)\s^2_2 + C^2(k)\s^2_3~,\eqno(3.44)$$
where the coefficient functions satisfy the relations \cite{ati}
$$ \eqalign{
AB ~=~ & -K(k)\left[E(k)-K(k)\right]~,\cr
BC~=~ & -K(k)\left[E(k)-{k'}^2K(k)\right]~,\cr
AC~=~ & -K(k)E(k)~.\cr}\eqno(3.45)$$
The parametrization (3.38) leads to the AH metric in the form (3.44) 
whose coefficient functions $(A,B,C)$ are given by a cyclic
permutation of those in eq.~(3.45) \cite{iro}. 
The K\"ahler potential of the AH metric was calculated in ref.~\cite{oli}.

In the limit $k\to 1$ (or, equivalently, $k'\to 0$), one has an 
asymptotic expansion
$$K(k)\approx -\log\, k'\left[ 1 +\fracmm{(k')^2}{4}\right] +\ldots~,
\eqno(3.46)$$
which suggests us to make a redefinition 
$$ k'=\sqrt{1-k^2}\approx 4\exp\left(\fracmm{1}{\g}\right)~,\eqno(3.47)$$
and describe the same limit at $\g\to 0^-$. After substituting eq.~(3.46) 
into eq.~(3.45) one finds that the AH metric becomes 
{\it exponentially close} to the Taub-NUT metric in
the form (3.40) subject to the additional relations 
$$ A^2\approx B^2\approx \fracmm{1+\g}{\g^2}~~,\quad C^2\approx 
\fracmm{1}{1+\g}~~,\eqno(3.48)$$
and with the negative mass parameter $M=-\ha$ \cite{ati}. In terms of
the general hyper-K\"ahler potential (3.2) whose `Taub-NUT' parameter 
$\l$ is fixed, we are left with the one-parameter family of the hyper-K\"ahler
metrics, all having the same perturbative behaviour. In this context, 
the unique solution given by the AH metric follows from a calculation 
of the one-instanton contribution to the LEEA, which unambigously 
determines the last parameter \cite{dkmtv} (see subsect.~3.3. for 
another argument). It also implies that $\abs{\g}^{-1}$ is
proportional to the one-instanton action.

The extra $U(1)$ symmetry of the Taub-NUT metric (when compared to the AH
metric) is the direct consequence of the relation $A^2=B^2$ arising in the 
asymptotic limit described by eq.~(3.48). The vicinity of $k'\approx
0^+$ describes the region of the hypermultiplet moduli space where 
quantum perturbation theory applies, with the exponentially small 
(nonholomorphic) AH corrections to the Taub-NUT metric being
interpreted as the instanton/anti-instanton contributions. Those 
nonperturbative corrections are supposed to be related to the existence of 
BPS monopoles in the underlying non-abelian N=2 field theory. The AH 
metric, as the metric in the hypermultiplet quantum moduli space, was 
proposed by Seiberg and Witten \cite{sw3} in the context of 3d, N=4 
supersymmetric gauge field theories, where it can be related to the 
(Seiberg-Witten) gauge LEEA \cite{sw} via the c-map in three spacetime 
dimensions (sect.~1).

From the N=2 PSS viewpoint, the transition from the perturbative 
hypermultiplet LEEA to the nonperturbative one thus corresponds to the 
transition from the $O(2)$ holomorphic line bundle associated with the 
standard N=2 tensor multiplet to the $O(4)$ holomorphic line bundle 
associated with the N=2 projective $O(4)$ multiplet. The two holomorphic
bundles are topologically different: with respect to the standard 
covering of $CP(1)$ by two open affine sets, the $O(2)$ bundle has the 
transition functions $\x^{-2}$, whereas the $O(4)$ bundle has the 
transition functions  $\x^{-4}$. In general, the variable $Q_{(2j)}$ is 
the coordinate in the fiber of the $O(2j)$ line bundle.

\subsection{Atiyah-Hitchin metric and elliptic curve}

The {\it quadratic} dependence of $Q_{(2)}$ upon $\x$ in eqs.~(B.2) and
(B.10) allows us to interpret  $Q_{(2)}(\x)$ as a holomorphic (of degree 2)
section of PSS, fibered by the superfields $(\c,g)$ and topologically 
equivalent to the Riemann sphere described by an algebraic equation 
$y^2=\c-i\x g+\x^2\Bar{\c}$ with $y^2\equiv\left.Q_{(2)}\right|$. 
Similarly, the {\it quartic}
dependence of $Q_{(4)}$ upon $\x$ in eqs.~(B.2) and (3.26) allows
us to interpret  $Q_{(4)}(\x)$ as a holomorphic (of degree 4) section of 
PSS, fibered by the superfields $(\c,W,V)$ and topologically
equivalent to the {\it elliptic curve} $\S_{\rm Hyper.}$ defined by 
an algebraic equation
$$ y^2(\x)=\c+\x W+\x^2 V-\x^3\Bar{W}+\x^4\Bar{\c}~,\eqno(3.49)$$
where $y^2\equiv\left.Q_{(4)}\right|$. The non-perturbative hypermultiplet 
LEEA can, therefore, be encoded in terms of the genus-one Riemann
surface $\S_{\rm Hyper.}$ \cite{prep}. This result is quite similar to the
famous Seiberg-Witten description \cite{sw} of the exact LEEA in the $SU(2)$ 
N=2 super-Yang-Mills theory, in terms of the elliptic curve $\S_{\rm SW}$.

The twistor construction of the AH metric \cite{ati} is known to be 
closely related to the {\it spectral} curve $\S_{\rm H}$ \cite{hur}. 
The elliptic curve $\S_{\rm H}$ naturally arises in the uniformization 
process of the algebraic curve (3.39) in the Donaldson description of the AH 
space \cite{bakas}. This actually provides enough evidence to identify 
$\S_{\rm H}$ with $\S_{\rm Hyper.}$ \cite{prep}. 

The defining equation (3.49) can be put into the normal (Hurtubise)
form \cite{ati}
$$ \tilde{y}^2(\tilde{\x})= 
K^2(k)\tilde{\x}\left[ kk'(\tilde{\x}^2-1)+(k^2-{k'}^2)\tilde{\x}
\right]~.\eqno(3.50)$$
Eq.~(3.50) is simply related to another standard
(Weierstrass) form, $y^2=4x^3-g_2x-g_3$ \cite{fkra}. Therefore, in
accordance with refs.~\cite{ati,fkra}, the real period $\o$ of $\S_{\rm H}$ is
$$\o\equiv 4k_1~,\quad {\rm where}\quad 4k_1^2=kk'K^2(k)~,\eqno(3.51)$$
whereas the complex period `matrix' of $\S_{\rm H}$ is given by
$$\t=\fracmm{iK(k')}{K(k)}~.\eqno(3.52)$$

The normal form (3.50) is related to that of eq.~(3.49) by the 
projective $SU(2)$ transformation ({\it cf.} eqs.~(B.7) and (B.9))
$$\x=\fracmm{\bar{a}\tilde{\x}-\bar{b}}{b\tilde{\x}+a}~,\quad 
y=\fracmm{\tilde{y}}{(b\tilde{\x}+a)^2}~,\quad 
\abs{a}^2+\abs{b}^2=1~,\eqno(3.53)$$
whose complex parameters $(a,b)$ are functions of the Euler `angles' 
$(\vq,\j,\vf)$  (see eq.~(3.38) in subsect.~3.2) \cite{bakas},
$$\eqalign{
a~=~& e^{\frac{i}{2}\vf}\left[ \sqrt{\fracmm{1-k}{2}}
\sin\fracmm{\vq}{2}e^{-\frac{i}{2}\j}
-i\sqrt{\fracmm{1+k}{2}}\cos\fracmm{\vq}{2}e^{\frac{i}{2}\j}\right]~,\cr
b~=~& e^{\frac{i}{2}\vf}\left[ -\sqrt{\fracmm{1+k}{2}}
\sin\fracmm{\vq}{2}e^{-\frac{i}{2}\j}
-i\sqrt{\fracmm{1-k}{2}}\cos\fracmm{\vq}{2}e^{\frac{i}{2}\j}\right]~.\cr}
\eqno(3.54)$$
Eq.~(3.54) implies another parametrization of the quartic (3.49) \cite{bakas},
$$\eqalign{
\F~=~&-\fracmm{1}{4}K^2(k)e^{-2i\vf}\sin^2\vq \left(1+{k'}^2
\sinh^2\n\right)~,\cr
H~=~& -\fracmm{1}{2}K^2(k)\sin(2\vq)e^{-i\vf}\left(1+{k'}^2
\cos\j\tan\vq\sinh\n\right)~,\cr
V~=~& +\fracmm{1}{2}K^2(k)\left[2-{k'}^2+3\sin^2\vq\left({k'}^2
\cos^2\j-1\right)\right]~.\cr}\eqno(3.55)$$
where $\n\equiv \log\left(\tan\fracmm{\vq}{2}\right)+i\j$. 
The parametrization (3.55) is closely related to that of eq.~(3.38), 
by a cyclic permutation of the coefficient functions $(A,B,C)$ in eq.~(3.44).

At generic values of the AH modulus $k$, $0<k<1$, the roots of the 
Weierstrass form are all different from each other, while they all lie 
on the real axis, say, at $e_3<e_2<e_1<\infty=(e_4)$. Accordingly, the 
branch cuts are running from $e_3$ to $e_2$ and from $e_1$ to $\infty$.
 The $C_r$ integration contour in the PSS formulation of the exact 
hypermultiplet LEEA in eq.~(3.32) can now be interpreted as the 
contour integral {\it over the non-contractible $\a$-cycle of the 
elliptic curve} $\S_{\rm H}$ \cite{bakas}, again in very close analogy
to the standard writing of the Seiberg-Witten solution \cite{sw} 
in terms of the abelian differential $\l_{\rm SW}$ integrated over the 
periods of $\S_{\rm SW}$. 

The perturbative (Taub-NUT) limit $k\to 1$ corresponds to the
situation when $e_2\to e_1$, so that the $\b$-cycle of $\S_{\rm H}$ 
degenerates. The curve (3.50) then asymptotically approaches a complex line, 
$\tilde{y}\sim \pm K\tilde{\x}$. Another limit, $k\to 0$, leads to a 
coordinate {\it bolt-type singularity} of the AH metric in the standard 
parametrization (3.44) \cite{ati}. In the context of monopole
physics, $k\to 0$ corresponds to the coincidence limit of two centered 
monopoles. In the context of the hypermultiplet LEEA, 
$k\to 0$ implies $e_2\to e_3$, so that the $\a$-cycle of $\S_{\rm H}$ 
degenerates, as well as the whole hypermultiplet 
action associated with eq.~(3.32). The two limits, $k\to 1$ and 
$k\to 0$, are related by the modular transformation exchanging $k$ 
with $k'$, and $\a$-cycle with $\b$-cycle \cite{bakas}. 

The AH metric is known to be {\it only} regular and complete four-dimensional
hyper-K\"ahler metric with the purely rotational $SO(3)$ isometry
\cite{ati,klnt,gm}. Being regular means the absence of singularities, 
while completeness means that every curve of finite length in the 
hyper-K\"ahler manifold under consideration has a limiting point.  
It is worth mentioning that only regular (i.e. globally well-defined) 
hyper-K\"ahler metrics can be interpreted as the metrics governing the 
hypermultiplet LEEA. Since the most general four-dimensional 
$su(2)_R$-invariant hyper-K\"ahler metrics in HSS are given by the 
two-parametric family described by the hyper-K\"ahler potential (3.2), 
it is important to find a simple way by which one can distinguish the 
regular AH metric amongst them. This problem is also present in the 
generalized Legendre transform construction of hyper-K\"ahler metrics. 
The underlying elliptic curve provides a nice interpretation to this problem 
\cite{prep}. The most general integration contour $C_r$ in the PSS 
construction (3.32) is described by eq.~(3.30). As is clear from the results 
of this subsection, any non-trivial contour $C_r$ can be equally interpreted 
as a linear conbination, $C_r=n_1\a +n_2\b$, of the fundamental cycles, $\a$ 
and $\b$, of the underlying elliptic curve  $S_{\rm H}$, with some integral 
coefficients, $n_1$ and $n_2$. An integration over the $\b$-cycle is known to 
yield a bolt-type singularity \cite{ati}, and it thus has be excluded. The 
contour integration over the $\a$-cycle yields the regular AH solution. The 
value  $n_1$ of the `winding' number is obvioulsy not relevant for regularity 
of the metric.~\footnote{The $D_k$ gravitational instanton metrics are not 
regular \cite{chka,dancer,chal2}, and they only possess $U(1)$ \newline
${~~~~~}$ rotational isometry.} A generalization of the `regular' 1-cycle to 
the higher-genus spectral curve associated with the charge-$n$ (centered) 
monopole space (subsect.~4.2) just leads to the Ercolani-Sinha constraints 
\cite{ersin} mentioned in the Introduction.

\section{More hypermultiplets and larger gauge groups}

If the underlying N=2 supersymmetric gauge field theory has a larger
(of rank $r>1$) simple gauge group, such as $SU(n+1)$, $n>1$, there
may be more (magnetically charged) massive hypermultiplets $q^{A+}$,
$A=1,2,\ldots,2n$, in the LEEA, either in the Coulomb branch or in the 
Higgs branch. Their exact self-interaction is described by the 
hyper-K\"ahler NLSM governed by a hyper-K\"ahler potential 
$\ck^{(+4)}(q^{A+};u^{\pm}_i)$ in HSS,
$$ -\Lag_{LEEA}=q^+_AD^{++}q^{A+} +\ck^{(+4)}(q^{A+};u^{\pm}_i)~.\eqno(4.1)$$

The non-anomalous $SU(2)_R$ symmetry further implies that the
hyper-K\"ahler potential $\ck^{(+4)}$ should be independent upon harmonics. 
Therefore, the analytic function  $\ck^{(+4)}$ is given by a {\it real
quartic polynomial} of the analytic HSS superfields $q^{A+}$,
$$\ck^{(+4)}(q)=P^{(+4)}(q)
\equiv \l_{ABCD}q^{A+}q^{B+}q^{C+}q^{D+}~,\eqno(4.2)$$
whose coefficients $\l_{(ABCD)}$ are totally symmetric and are subject
to the reality condition, $\sbar{P}{}^{(+4)}=P^{(+4)}$. 
In a bit more explicit notation, $q^{A+}=(q^+_M,-\sbar{q}{}^+_M)$, we have
$$P^{(+4)}\equiv \l_{MN,\bar{P}\bar{Q}}
q^+_Mq^+_N\sbar{q}{}^+_{\bar{P}}\sbar{q}{}^+_{\bar{Q}}+
\bar{\l}_{NMPQ}q^+_Mq^+_Nq^+_Pq^+_Q +
\bar{\l}_{NMP\bar{Q}}q^+_Mq^+_Nq^+_P\sbar{q}{}^+_{\bar{Q}} 
+{\rm h.c.}~,\eqno(4.3)$$
with some constants $\l_{(MN),(\bar{P}\bar{Q})}$,
$\bar{\l}_{(NMPQ)}$ and $\bar{\l}_{(NMP)\bar{Q}}$. 

Not all of the constants are really significant since the kinetic
terms in eq.~(4.1) have the manifest global $Sp(n)$ symmetry under the 
transformations of $q^{A+}$. Hence, the space of all $SU(2)_R$-invariant 
hyper-K\"ahler metrics in 4n real dimensions is parametrized by the quotient
$$ \ct_m=\{P^{(+4)}(\l)\}/Sp(n)~~.\eqno(4.4)$$
It may not be accidental that the $Sp(n)$ factor in eq.~(4.4) coincides with 
the maximal $Sp(n)$ holonomy group of hyper-K\"ahler manifolds in $4n$ real 
dimensions.

\subsection{Exact hypermultiplet LEEA in the Coulomb branch}

In the Coulomb branch of the N=2 super-Yang-Mills theory with the 
$G=SU(n+1)$ gauge group, the gauge symmetry $G$ is (generically) 
spontaneously broken to its maximal Cartan subgroup $H=U(1)^n$, 
due to a non-vanishing 
vacuum expectation value, $\VEV{\f}\neq 0$, of the adjoint Higgs field. Since
$\p_2[SU(n+1)/U(1)^n]=\p_1[U(1)^n]={\bf Z}^n$, one expects ${\rm rank}(G)=n$
{\it different} types of magnetic monopoles associated to each of the Cartan 
generators and belonging to hypermultiplets. The corresponding classical 
solitonic solutions can be obtained via embedding the known $SU(2)$ 
solutions (see Appendix D) along the simple root directions in $G$ 
(see ref.~\cite{lwp}, or a recent review \cite{wein}). As regards
the LEEA of the hypermultiplets corresponding to {\it different} simple 
roots, it should possess the tri-holomorphic (or translational) $U(1)^n$ 
isometry, in addition to the rotational $SU(2)_R$ isometry discussed
above. In this subsection, we present a very simple derivation of the 
corresponding hyper-K\"ahler potential in HSS, and give the associated
hyper-K\"ahler metric in components.

First, let's recall some basic facts about Lie algebras and monopoles 
\cite{gno}. The generators of a rank-$r$ Lie algebra~\footnote{Most 
of our considerations apply to {\it any} simple Lie group $G$. We use 
$SU(n+1)$ for definiteness.} can be naturally divided into $r$
commuting Cartan generators $\{H_i\}$, and the raising and lowering operators,
$E_{\vec{\a}}$ and $E_{-\vec{\a}}$, for each of the $r$-component root
vectors $\{\vec{\a}\}$,
$$ \[ H_i,H_j\]=0~,\quad \[ E_{\vec{\a}},\vec{H}\]=\vec{\a}E_{\vec{\a}}~,
\quad \[E_{\vec{\a}},E_{-\vec{\a}}\]=\vec{\a}\cdot\vec{H}~.\eqno(4.5)$$
We assume the standard normalization, $\tr(H_iH_j)=\d_{ij}$. The root
vectors define a root lattice, while any root vector can be decomposed
with respect to a basis of {\it simple} (fundamental) roots  
$\{\vec{\a}_i\}$ with non-negative integral coefficients,
$$ \vec{\a}=\sum^r_{i=1}n_i\vec{\a}_i~.\eqno(4.6)$$
The simple root basis is not unique, with all choices being related by the
Weyl group transformations. The {\it dual} root lattice is defined by 
$$ \vec{\a}^*=\sum^r_{j=1}n_j^*\vec{\a}^*_j~,\quad{\rm where}\quad
\vec{\a}^*_j=\vec{\a}_j/\vec{\a}_j^2~.\eqno(4.7)$$
Given $G=SU(n+1)$, one has $\vec{\a}_i\cdot\vec{\a}_i>0$ and
$\vec{\a}_i\cdot\vec{\a}_j\leq 0$ for $i\neq j$; while $r=n$.

The vacuum expectation value of the Higgs field can always be assigned
to the Cartan subalgebra,
$$\VEV{\f}=\vec{h}\cdot\vec{H}~.\eqno(4.8)$$  
If $\vec{h}\in H$ has non-vanishing products with all roots of $G$, 
as we assume here, the group $G$ is maximally broken to its Cartan
subgroup. The unique set of simple roots is naturally distinguished 
by the condition
$$ \vec{h}\cdot \vec{\a}_i > 0~,\quad{\rm where}\quad
 1\leq i\leq r~.\eqno(4.9)$$

A magnetic charge is characterized by another vector $\vec{g}\in H$.
The topological (Dirac) quantization condition on the magnetic charge
implies that $\vec{g}$ belongs to the dual lattice ({\it cf.} Appendix D),
$$ \vec{g}=\fracmm{4\p}{e}\sum_{j=1}^r k_j\vec{\a}^*_j\equiv
g\sum_{j=1}^r k_j\vec{\a}^*_j\equiv
\sum_{j=1}^r k_j\vec{g}_j~,\eqno(4.10)$$
where $\{k_j\}$ are topological charges belonging to the
integral homotopy ${\bf Z}^{r}$, and $g=4\p/e$ is the unit of magnetic charge.
The general BPS mass formula reads 
$$ M=\abs{\vec{h}\cdot\vec{g}}~,\eqno(4.11)$$
so that the mass spectrum of the fundamental monopoles (of unit
charge) is given by
$$ m_j=\abs{\vec{h}\cdot\vec{g}_j}~,\quad j=1,2,\ldots,r~.
\eqno(4.12)$$
The long-range force between two different monopoles $i$ and $j$ obeys
the standard (inverse radius squared) law, whose strength is given
by the inner product of magnetic charges, $\vec{g}_i\cdot\vec{g}_j/4\p$
\cite{wein,gno}. The classical solitonic solutions describing
$SU(n+1)$ monopoles can be constructed, in principle, by the ADHMN method
\cite{adhmn}, whose (Nahm) data is, however, highly constrained. It makes a 
direct construction of the multi-monopole moduli space metrics via the 
so-called `moduli space approximation' \cite{manton} to be highly non-trivial.

Each of $n$ (translational) commuting $U(1)$ isometries can be realized  by 
rigid phase rotations of a single $q$-hypermultiplet in eq.~(4.3) 
({\it cf.} eq.~(2.51)),
$$ q^+_m \to e^{i\a\low{m}}q^+_m\quad {\rm and}\quad  
\sbar{q}{}^+_m \to e^{-i\a\low{m}}\sbar{q}{}^+_m~,\quad{\rm where}\quad
m=1,2,\ldots,n~.\eqno(4.13)$$ 
This implies that the most general quartic polynomial of eq.~(4.3) reduces
in this case to a merely {\it quadratic} real function of the
invariants
$$iL^{++}_m\equiv\,\sbar{q}{}^+_m q^+_m~,\eqno(4.14)$$
which reads
$$ P^{(+4)}=\ha\sum_{p,q=1}^n \l_{pq}L^{++}_pL^{++}_q\equiv
\ha\sum_{j=1}^n \l_j(\sbar{q}{}^+_j q^+_j)^2 +\sum_{i<j}^n   \l_{ij}
(\sbar{q}{}^+_i q^+_i)(\sbar{q}{}^+_j q^+_j)~.\eqno(4.15)$$
A perturbative calculation of the HSS multi-hypermultiplet one-loop
diagram (Fig.~1) in the N=2 dual $U(1)^n$ gauge theory also yields \cite{ikz}
$$ P^{(+4)} \propto \sum_{i,j=1}^n 
(\sbar{q}{}^+_i \vec{g}_i\,q^+_i)\cdot(\sbar{q}{}^+_j\vec{g}_j\, q^+_j)~.
\eqno(4.16)$$
We thus conclude, from the symmetry considerations alone, that the 
hypermultiplet LEEA decomposes into the sum of the {\it single} 
hypermultiplet self-interactions and their {\it pairwise} interactions.

Moreover, the equations of motion for the $q$-superfields, in the
presence of any $SU(2)_R$-invariant self-interaction $\ck^{(+4)}(q)$, 
imply the conservation laws $D^{++}L^{++}_m=0$ for the composite HSS
superfields (4.14) subject to the reality condition $\sbar{L^{++}_m}=L_m^{++}$.
In other words, $L^{++}_m$ can be recognized as the N=2 {\it tensor} 
multiplets ({\it cf.} sect.~3). This confirms the claim \cite{chal}
that the exact metric on the moduli space of $n$ distinct monopoles in 
a spontaneously broken $SU(n+1)$ pure gauge theory can be obtained by the 
generalized Legendre transform in terms of sections of the $O(2)$ holomorphic 
bundle only. In our terms, this result comes after projecting the HSS
action (4.15) into PSS by using the N=2 duality transformation (4.14). 
The equivalent PSS action is given by a sum of the non-improved 
and improved terms \cite{chal},
$$  \oint G = \sum_{j=1}^n m_j\oint_{C_0} \fracmm{Q_{j(2)}^2}{2\x} + 
\sum^n_{i<j}\fracm{1}{2\p}(\vec{g}_i\cdot\vec{g}_j)
\oint_{C_r} (Q_{i(2)}-Q_{j(2)})\left\{ \ln
(Q_{i(2)}-Q_{j(2)})-1\right\}~,\eqno(4.17)$$
in agreement with eqs.~(4.15) and (4.16).

The generalized Legendre transform (Appendix B) of the N=2 PSS action 
defined by eqs.~(B.3) and (4.17) yields the N=2 NLSM whose hyper-K\"ahler 
metric is given by \cite{lwp,chal}
$$ ds^2_{\rm mTN}=M_{ij}d\vec{r}_i\cdot d\vec{r}_j +
\fracmm{g^4}{(4\p)^2}(M^{-1})_{ij}(d\x_i+\vec{W}_{ik}\cdot d\vec{r}_k)
(d\x_j+\vec{W}_{jl}\cdot d\vec{r}_l)~,\eqno(4.18)$$
where
$$ M_{ij}=\left\{ \begin{array}{cc} m_i -\sum_{k\neq i}
\fracmm{\vec{g}_i\cdot\vec{g}_k}{4\p r_{ik}}~~, & i=j~, \\
\fracmm{\vec{g}_i\cdot\vec{g}_j}{4\p r_{ij}}~~, & i\neq j~, 
\end{array} \right. \eqno(4.19)$$
and
$$ \vec{W}_{ij}=\left\{ \begin{array}{cc} -\sum_{k\neq i}
\vec{\a}^*_i\cdot\vec{\a}^*_k\,\vec{w}_{ik}~~, & i=j~, \\
\vec{\a}^*_i\cdot\vec{\a}^*_j\,\vec{w}_{ij}~~, & i\neq j~.
\end{array} \right. \eqno(4.20)$$
The quantity $\vec{w}_{ij}$ is the Dirac potential from the $j$-th
monopole evaluated at the position $\vec{r}_i$ where the $i$-th Dirac monopole
is located. The function  $\vec{w}_{ij}(\vec{r}_i-\vec{r}_j)$ satisfies an 
equation
$$ \vec{\de}_i\times\vec{w}_{ij}(\vec{r}_i-\vec{r}_j)=
\fracmm{-(\vec{r}_i-\vec{r}_j)}{r_{ij}^3}~~.\eqno(4.21)$$

The metric (4.18)--(4.21) was first identified by Lee, Weinberg and Yi (LWY) 
\cite{lwp} as the {\it asymptotic} metric in the multi-monopole moduli space 
for the $n$  {\it distinct} fundamental monopoles corresponding to 
all different simple (or fundamental) roots. Because of its origin,
the LWY metric may also be called the {\it multi-dimensional Taub-NUT} 
metric. Since the inner products of the dual 
fundamental roots in the case of $SU(n+1)$ can only be either zero or 
negative, the hyper-K\"ahler metric (4.18)--(4.21) appears to be regular and 
geodesically complete \cite{lwp}. Subsequently, it was further argued
by the same authors \cite{lwp} that the metric found may actually be 
{\it exact} in this case (see ref.~\cite{wein} too). As was later 
demonstrated by Chalmers \cite{chal}, by the use of the generalized Legendre  
transform based on eq.~(4.17), the LWY metric is unique indeed. Our
HSS approach greatly simplifies the proof of this statement.

If one deals with $n>1$ similar monopoles (or hypermultiplets) labelled by the
same root (say, $\vec{g}_1$), the multi-dimensional Taub-NUT metric develops 
a singularity in the `core' region at $r=\vec{g}{}^2_1/(4\p m)$, where the
similar monopoles approach each other \cite{lwp}. It means that the asymptotic
LWY metric cannot be exact in this case. This situation is apparently similar 
to the single charge-2 case described by the AH metric whose asymptotic 
behaviour is  given by the Taub-NUT metric with the negative mass parameter 
(sect.~3). 

The $SU(2)_R$ invariant family of hyper-K\"ahler potentials in 
eq.~(4.2) describes the exact metrics on the multi-monopole moduli space 
if there are no more than two similar monopoles belonging to the same root,
$k_i\leq 2$. In other words, this situation corresponds to `gluing' together 
$n$ Atiyah-Hitchin or Taub-NUT metrics, each being associated with a simple 
root.~\footnote{The case of $k_i>2$ is discussed in the next subsect.~4.2.} 

A generic real quartic polynomial $P^{(+4)}(q)$ may have no translational 
isometries at all. The problem of extracting the explicit hyper-K\"ahler 
metric associated with a hyper-K\"ahler potential (4.2) essentially amounts to 
 determination of the set of composite analytic HSS superfields of the 
second- and forth-order in $q^+$, which would allow one to make an N=2 
supersymmetric reduction from HSS to PSS, and thus to get rid of the 
auxiliary fields. One such variable is obviously given by the hyper-K\"ahler 
potential $P^{(+4)}(q)$ itself, because the conservation law (3.3) is valid 
for {\it any} number of the FS hypermultiplets subject to their equations 
of motion (3.4). This gives rise to the $O(4)$ projective multiplet 
$L^{++++}=P^{(+4)}(q)$ as one of the `good' variables. If $P^{(+4)}(q)$ can 
be represented as a quadratic polynomial squared, the role of $L^{++++}$ is 
replaced by an N=2 tensor multiplet.  One can easily imagine mixed 
situations, where dualizing the HSS potential $P^{(+4)}(q)$ can be performed 
by using both $O(2)$ {\it and\,} $O(4)$ projective N=2 multiplets. Some of 
those `mixed' PSS actions were considered in ref.~\cite{chal}. 

\subsection{Non-abelian LEEA, rational maps, and spectral curve}

It should be emphasized that in the HSS-based approach we may
easily treat {\it all} types of the gauge symmetry breaking. We only 
discussed above the {\it maximal} symmetry breaking of
the gauge group $G$ down to its Cartan subgroup. More general patterns 
of the gauge symmetry breaking are possible, with the unbroken subgroup 
being of the form $K\times U(1)^{r-k}$, where $K$ is a non-abelian 
(of rank $k<r$) subgroup of $G$. It happens when the vector $\vec{h}\in H$, 
defining the vacuum expectation value of the Higgs field in eq.~(4.8),
is orthogonal to the simple roots of $K$. The unbroken symmetry should be 
simply imposed on the hypermultiplet LEEA as its isometry. It just forces the
 hyper-K\"ahler potential (4.3) to be a function of the 
$K\times U(1)^{r-k}$ invariants only, which considerably restricts the form 
of the quartic (4.3). In addition, the magnetic charge $\vec{g}$ is supposed 
to commute with the unbroken symmetry, which implies \cite{wein}
$$ \vec{g}\cdot \vec{\g}_i=0\eqno(4.22)$$
for all simple roots $\vec{\g}_i$ of $K$, $i=1,2,\ldots,k$. 

In the case of {\it massless} monopoles and hypermultiplets, some of
the diagonal coefficients  of the quartic (4.3) become infinite. It
does not, however, imply that the HSS description becomes invalid. Although
it means that some rescalings are necessary. This is particularly clear 
in the PSS description, in terms of the N=2 tensor multipets, where eq.~(4.17)
apparently continues to be valid, even if some of the masses $m_i$ vanish.

Given the existence of the elliptic curve behind the AH metric associated with
magnetic charge 2, it is not very surprising that there also exist the
Riemann surface $\S_g$ of genus  $g=(n-1)$ in the case 
of magnetic charge $n$ (or, equivalently, $n$ similar fundamental monopoles). 
The $\S_g$ is known as the {\it spectral curve} in the twistor theory of 
monopoles \cite{hitchin}. We now briefly describe a relation between the 
PSS and HSS constructions of the multi-monopole moduli space metrics and the 
spectral curve. The good starting point is the Donaldson 
classification \cite{don} of the charge-$n$ (BPS) monopoles in the $SU(2)$ 
Yang-Mills-Higgs system (Appendix D). It is just opposite to the situation 
discussed in the preceeding subsect.~4.1, since now all monopoles are supposed
to be similar, being assigned to a single root of $G$.

The Donaldson classification \cite{don} makes use of the 
{\it rational maps} defined by
$$ S(z)=\fracmm{p(z)}{q(z)}\equiv\fracmm{\sum_{j=0}^{n-1} a_jz^j}{z^n +
\sum_{j=0}^{n-1} b_{n-j}z^j}\eqno(4.23)$$
in $R_n(z)$. It is assumed that the roots $\b_j$ of $q(z)$ are all different,
 while the residues $p(\b_j)\neq 0$ for all $j$. According to 
ref.~\cite{ati}, there exist a closed 2-form,
$$ \O=\sum^n_{j=1}d\b_j \wedge d\ln p(\b_j)~,\eqno(4.24)$$
that is symmetric under any exchange of $\b_i$, symplectic in $R_n(z)$, and
holomorphic, being an $O(2)$-section over $CP(1)$. The Donaldson theorem 
\cite{don} claims the existence of a {\it one-to-one} correspondence between 
the (universal cover of) charge-$n$ monopole configurations and the rational 
maps $S(z)$. For example, the $O(2)$-section $\O$ can be (locally) written 
down as
$$ \O= (J^{(2)}+iJ^{(3)})+2\x J^{(1)}-\x^2(J^{(2)}-iJ^{(3)})~,\eqno(4.25)$$
where $(J^{(1)},J^{(2)},J^{(3)})$ is the hyper-K\"ahler structure \cite{ati}.

It is worth mentioning that the input provided by the rational map (4.23) 
naively has $4n$ real parameters given by the real and imaginary parts of the 
complex coefficients $(a_i,b_j)$. Since the function $S(z)$ transforms as 
the (chiral) 2d conformal field \cite{book} under the transformations   
$$ a_i\to \l^i a_i~,\quad b_i\to \l^{-i}b_i~,\eqno(4.26)$$
with the complex parameter $\l$, the map $S(z)$ is actually 
dependent upon $(4n-2)$ real parameters. The gauge symmetry (4.26)  
can be fixed, for example, by choosing a gauge $b_{n-1}=0$. The equation
$$\D_n(p,q)\equiv \prod_{j=1}^n p(\b_j)=1~,\eqno(4.27)$$
where $\D_n(p,q)$ is known as the {\it resultant} of the map $S(z)$, defines
the algebraic manifold $\Tilde{\cm}_n$ of real dimension $4(n-1)$. 
The resultant also has a cyclic symmetry under the transformations
$$\G_n:\qquad a_i \to qa_i~,\quad {\rm where}\quad q^n=1~.\eqno(4.28)$$
As is explained in ref.~\cite{don}, it is the quotient space 
$\Tilde{\cm}_n/\G_n$ that is supposed to be identified with the charge-$n$ 
(centered) monopole moduli space ({\it cf.} eq.~(3.39) for the AH space). 
Setting $\D_n=1$ and $b_{n-1}=0$ means choosing the overall phase and location
of the $n$-monopole to be zero \cite{ch}. It is now easy to verify that the
Donaldson description of the AH space in eq.~(3.39) follows from eq.~(4.27) 
in the case of $n=2$.

In the general case, all $\{\b_i,p(\b_j)\}$ are the good complex (Donaldson) 
coordinates on the multi-monopole moduli space, while $b_j$ can be
recognized as the $O(2j)$ section over $CP(1)$, $b_j(\x)=Q_{(2j)}(\x)$, 
that is real in the sence of eq.~(2.33).  The (Hitchin) spectral curve
$\S_g$ of genus $g=n-1$ can now be written down in the form
$$ y^n + \sum^n_{j=1} \x^{n-j}Q_{(2j)}(\x)=0~.\eqno(4.29)$$  

The Riemann surface (4.29) has the full information about a charge-$n$ 
monopole. The equivalent PSS data $(G,C)$ in the form of  eq.~(B.3) and 
the related HSS potential $\ck^{(+4)}$ in eq.~(2.59) can, therefore, be 
completely encoded in terms of the spectral curve (4.29). 
In particular, the corresponding PSS potential was recently found by 
Houghton \cite{hou}, and it leads to the N=2 PSS NLSM action dictated by 
$$ \fracmm{1}{2\p i}\oint G = \oint_C y - \fracmm{1}{2\p i}
\oint_O \fracmm{y^2}{\x}~,\eqno(4.30)$$
where $C$ is a non-trivial 1-cycle on $\S_g$ (see, e.g., ref.~\cite{ersin} for
details), and $O=\sum^n_{j=1}0_j$ where $0_j$ encircles the origin of the 
$j$-th sheet of $\S_g$.
The corresponding HSS potential $\ck^{(+4)}(q,u)$ is a {\it finite}
polynomial in $q$  or order $2n$, while it must be explicitly dependent upon
harmonics in the case of $n>2$. This means that the $SU(2)_R$ symmetry is
necessarily broken for $n>2$. The asymptotical behavior of the multi-monopole
 moduli space metric is described by the limit where the spectral curve $\S_g$ 
degenerates (with exponential accuracy) into the product of $g$ spheres
\cite{hou}. It also implies that $Q_{(2j)}\to Q_{(2)}^j$ that brings us back
to the preceeding subsect.~4.1. 

Since the HSS method can be considered as the very particular application 
of {\it flag} manifolds in the twistor approach to a construction of the
hyper-K\"ahler metrics on the multi-monopole moduli spaces, the HSS method
may also be useful for a construction of the classical monopole solutions 
themselves. For instance, a correspondence between the $SU(n+1)$ monopoles 
and the rational maps of the Riemann sphere into flag manifolds was
recently noticed in ref.~\cite{jarvis}, whereas the relevance of 
{\it harmonic} maps in this connection was recently emphasized in 
ref.~\cite{mhpa}. It would be interesting to explore their relations to HSS.

\section{Conclusion}

In this paper we reviewed many aspects of the hypermultiplet low-energy 
effective action, including the PSS and HSS technology. Our results are 
summarized in the Abstract. We conclude with a few remarks of general nature.

The 4d twistor approach has many similarities with the inverse
scattering method in the theory of (lower dimensional) 
{\it integrable systems}. It its turn, the integrable systems are
known to be closely connected to the hyper-K\"ahler geometry and the
theory of Riemann surfaces \cite{dwi}. In our approach to a
construction of the 4d hypermultiplet LEEA, it was the passage from 
the N=2 harmonic superspace (HSS) to the N=2 projective superspace 
(PSS) that provided the link of harmonic analysis to complex analysis and
thus allowed us to introduce the holomorphic quantities. The latter were
then interpreted in terms of the elliptic curve $\S_{\rm H}$. In addition,
the uniformization of the AH algebraic curve (3.39) describing 
the AH space is known to be closely related to the continual 3d 
{\it Toda equation} \cite{bakas} that, in its turn, naturally arises 
in the large $N$ limit of 2d conformal field theories \cite{book}. 

It is also remarkable that the very simple (quartic polynomial) `Ansatz' 
(4.2) for the hyper-K\"ahler potential in HSS provides the exhaustive 
description of the highly non-trivial class of the $SU(2)_R$-invariant
hyper-K\"ahler metrics that naturally generalize the standard 
(four-dimensional) Taub-NUT and Atiyah-Hitchin metrics to higher dimensions.
The related hypermultiplet LEEA has manifest N=2 supersymmetry and the 
$SU(2)_R$ internal symmetry. In particular, the N=2 NLSM in HSS with the
Atiyah-Hitchin metric has the natural holomorphic projection to N=2
PSS, where it is associated with the holomorphic $O(4)$ line bundle. Similar 
projections exist in the more general case of $n$ fundamental monopoles.

The non-perturbative (instanton) corrections to the hypermultiplet LEEA are 
dictated by the hidden (in 4d) spectral curve parametrizing the exact solution.
Since the Seiberg-Witten curve $\S_{\rm SW}$ is known to have the simple 
geometrical interpretation in M-theory, where it can be considered as the 
part of a magnetically charged five-brane worldvolume wrapped about 
$\S_{\rm SW}$ \cite{w}, it is conceivable that the hypermultiplet spectral
curve $\S_{\rm H}$ may have, perhaps, a similar geometrical interpretation 
which is presumably related to (Dirichlet) 6-branes in ten-dimensional 
spacetime \cite{rev2}.

\section*{Acknowledgements}

I am grateful to M.~Arai, I.~Bakas, S.~Cherkis, S.~Gates~Jr., K.~Ito,
O.~Lechtenfeld, T.~Nakatsu, N.~Ohta, N.~Sakai and P. Yi for useful discussions.
I also thank the referee for advising me about the references that somehow
escaped my attention or were improperly cited.    

\newpage

\section*{Appendix A: ABC of hyper-K\"ahler geometry}

A full account of complex geometry, including the hyper-K\"ahler one, 
is available in the mathematical literature, see e.g., the books 
\cite{difg,besse}. We follow here the presentation of ref.~\cite{haklr}
adapted for physicists.

Let's consider a manifold $\cm$ of real dimension $2k$, covered by 
a system of coordinate neighbourhoods (charts) $\{x^i\}$. The coordinates
in an intersection of two charts are supposed to be related by smooth 
and locally invertible {\it transition functions}.

A mixed second-rank tensor $J\du{i}{j}$ with real components is
called an {\it almost complex structure} if it satisfies the relation
$$ J\du{i}{j} J\du{j}{k}=- \d\du{i}{k}~.\eqno(A.1)$$
The manifold $\cm$ equipped with an almost complex structure is called 
an {\it almost complex manifold} of complex dimension $k$. The almost 
complex structure defines a multiplication by `$i$' on vectors in every
coordinate chart. The projectors 
$$ P_{\pm} =\frac{1}{2}(1\pm iJ) \eqno(A.2)$$
can be used to split any vector $V^i$ into two projections
$V^i_{\pm}$. In particular, the basis 1-forms $dx^j$  can be split into 
$$ \o^i_{\pm}= P\du{\pm j}{i}dx^j~.\eqno(A.3)$$

The almost complex structure is said to be {\it integrable} if
$$  (d\o_{\pm})^k_{\mp}= P\du{\mp i}{k}dP\du{\pm
  j}{i}dx^j=0~. \eqno(A.4)$$
In this case $J$ is called a {\it complex structure}, while $\cm$ is 
called a {\it complex manifold}. The complex manifold is 
characterized by {\it holomorphic} transition functions.
Eq.~(A.4) is equivalent to the vanishing torsion on $\cm$,
$$ N\du{ij}{k}\equiv J\du{[i|}{n}\pa_n J\du{|j]}{k} +
J\du{n}{k}\pa_{[j}J\du{i]}{n}=0~,\eqno(A.5)$$
where $N\du{ij}{k}$ is known as Nijenhuis tensor.

After introducing holomorphic and antiholomorphic coordinates $z^i$ 
and $\bar{z}^{\bar{i}}$ in $\cm$, eq.~(A.4) or (A.5) allows one to put 
the complex structure $J$ into the {\it canonical} form
$$ J\du{j}{i}=\left( \begin{array}{cc} i\d^j_i & 0 \\
0 & -i\d^{\bar{j}}_{\bar{i}} \end{array} \right)~~.
\eqno(A.6)$$
The existence of a system of complex coordinate
charts in which the almost complex structure has the form
(A.6) is equivalent to the integrability condition (A.4) \cite{difg}. 

Given a Riemannian manifold $\cm$ with a metric
$g_{ij}$ and an (almost) complex structure $J\du{i}{j}$,
the invariance of the metric with respect to the 
complex structure means
$$ J\du{i}{k}J\du{j}{m}g_{km}=g_{ij}~,\quad {\rm 
  or,~~equivalently,} \quad J_{ij}\equiv 
g_{jk}J\du{i}{k}=-J_{ji}~~.\eqno(A.7)$$
The metric satisfying eq.~(A.7) is called {\it  hermitian}.

An (almost) complex manifold equipped with a hermitian
metric is called an (almost) {\it hermitian manifold}. 
The hermitian manifold thus possesses the {\it fundamental} 2-form 
$$ \o \equiv J_{ij}\, dx^i\wedge dx^j~.\eqno(A.8)$$
In the special coordinates (where the complex
structure $J$ is of the canonical form), the fundamental 2-form reads
$$ \o = 2ig_{i\bar{j}}\, dz^i\wedge
d\bar{z}^{\bar{j}}~~.\eqno(A.9)$$

An (almost) {\it K\"ahler manifold} is the (almost)
hermitian manifold with the {\it closed} fundamental form, 
$$ d\o=0\quad {\rm or,~~equivalently,} \quad 
J_{[ij,k]}=0~~.\eqno(A.10)$$
As is clear from eqs.~(A.9) and (A.10), the metric of
a K\"ahler manifold can be (locally) expressed in terms
of a {\it K\"ahler potential}. The conditions (A.5) and
(A.10) together are equivalent to the covariant constancy
of the complex structure,
$$ \de_iJ\du{j}{k}=0~~.\eqno(A.11)$$

An {\it (almost) quaternionic structure} is the set of three 
linearly independent (almost) complex structures $J\du{i}{(A)j}$,
$A=1,2,3$, satisfying the quaternionic algebra,
$$ J\du{i}{(A)k}J\du{k}{(B)j}=-\d^{AB}\d^j_i +
\ve^{ABC} J\du{i}{(C)j}~~.\eqno(A.12)$$
A manifold with the quaternionic structure is
called a {\it quaternionic manifold}. If a quaternionic manifold 
possesses a metric that is hermitian with respect to all three
 covariantly constant complex structures, the manifold is called 
{\it  hyper-K\"ahler}. In other words, a hyper-K\"ahler manifold 
is characterized by three linearly independent complex structures 
satisfying the quaternionic algebra, while there exists the coordinate
system for any given complex structure where it takes the 
canonical form (A.6). 

Any hyper-K\"ahler manifold is Ricci-flat. In four dimensions
the hyper-K\"ahler condition is, in fact, equivalent to the 
Ricci-flatness and K\"ahler conditions together. Imposing the 
Ricci-flatness condition on a K\"ahler potential $K$ --- see eq.~(2.8) --- 
results in the non-linear partial differential equation
$$ \det(K\du{i}{j})=1 \eqno(A.13)$$ 
known as the Monge-Amp\'ere equation \cite{hklr}.

Perhaps, the most elegant description of hyper-K\"ahler geometry is 
possible in terms of {\it holonomy}, whose generators are the components
of the Riemannian curvature. Given a Riemannian manifold of real 
dimension $4n$, it is hyper-K\"ahler if and only if its holonomy is a 
subgroup of $Sp(n)$. The case of $n=1$ is special since the most
general 4d holonomy group factorizes, $O(4)\cong SU(2)\times SU(2)$. 
Self-duality of the Riemann curvature means that the holonomy is a 
subgroup of $SU(2)$. Since $SU(2)\cong Sp(1)$, it is clear that a 
four-dimensional hyper-K\"ahler manifold is always (anti)self-dual 
and vice versa \cite{ati}.

\section*{Appendix B: N=2 projective superspace (PSS)}

The idea of N=2 projective superspace (PSS) naturally comes out of the
efforts to construct an N=2 supersymmetric self-interaction of the 
projective $O(n)$ multiplets satisfying the constraints (2.29) and 
(2.30) \cite{klr,ok}. Let's introduce a function $G(L_A;\x,\h)$,
whose arguments are given by some number $(A=1,2,\ldots)$ 
of the $O(n)$ projective multiplets (of any type with $n\geq 2$) 
and two projective (complex) $CP(1)$ coordinates, $\x$ and $\h$, and
constrain it by four linear differential equations in N=2 superspace, 
$$ \de_{\a}G\equiv (D^1_{\a}+\x D^2_{\a})G=0~,\quad
\D_{\dt{\a}}G\equiv (\Bar{D}^1_{\dt{\a}}+\h\Bar{D}^2_{\dt{\a}})G=0~.
\eqno(B.1)$$
It is straightforward to verify that a general solution to eq.~(B.1) reads
$$ G=G(Q_A(\x);\x)~,\quad \h=\x~,\quad 
Q_{(n)}(\x)\equiv \x_{i_1}\cdots \x_{i_n}
L^{i_1\cdots i_n}~,\quad \x_i\equiv (1,\x)~,\eqno(B.2)$$
in terms of an {\it arbitrary} function $G(Q_A;\x)$. Of course, 
in deriving eq.~(B.2) we have used the defining constraints (2.29).

Since the function $G$ does not depend upon a half of the Grassmann
coordinates of N=2 superspace by its definition (B.1), its integration over 
the rest of the N=2 superspace coordinates is invariant under N=2 
supersymmetry. This leads to the following N=2 invariant action \cite{klr,ok}:
$$ S[L_A] = \int d^4x\,\fracmm{1}{2\p i}\oint_C d\x\,(1+\x^2)^{-4}\Tilde{\de}^2
\Tilde{\D}^2G(Q_A,\x)+{\rm h.c.}~,\eqno(B.3)$$
where we have introduced the new superspace derivatives, 
$$ \Tilde{\de}_{\a}\equiv \x D^1_{\a}-D^2_{\a}~,\quad \Tilde{\D}_{\dt{\a}}
\equiv \x\Bar{D}^1_{\dt{\a}}-\Bar{D}_{\dt{\a}}^2~,\eqno(B.4)$$
in the directions orthogonal to the `vanishing' directions defined by 
eq.~(B.1). The integration contour $C$ in the complex $\x$-plane is supposed 
to make the action (B.3) non-trivial. The points $\x_{\pm}=\pm i$, 
where the linear independence of the derivatives (B.1) and (B.4) 
breaks down, should be outside of the contour $C$.

The form of the PSS action (B.3) is universal in the sense that it applies to 
any set of the projecive  multiplets $L_{(n)}$ entering the
action via the corresponding function $Q_{(n)}(\x)$ defined by eq.~(B.2),
while the whole action is governed by a single holomorphic potential $G$.
This construction is easily generalizable to the case of the so-called
{\it relaxed} hypermultiplets.  For example, in the case
of the relaxed Howe-Stelle-Townsend hypermultiplet defined by the N=2
superspace constraints \cite{hst}
$$ D_{\a}^{(i}L^{jk)}=D_{\a l}L^{ijkl}~,\quad D_{\a}^{(i}L^{jklm)}=0~,
\eqno(B.5)$$
and their conjugates, one should merely substitute $Q_{(2)}$ by \cite{oldket}
$$ Q_{(2),{\rm rel.}}=Q_{(2)}-\fracmm{5}{4}\fracmm{\pa Q_{(4)}}{\pa\x}~.
\eqno(B.6)$$

Some comments are in order. In the odd case of $n=2p+1$, the
conjugated superfields $\Bar{L}^{i_1\cdots i_n}$ may enter the action (B.3)
via the corresponding polynomial $\Bar{Q}_{(2p+1)}(\x)$. However, this makes
the potential $G$ to be nonholomorphic, so that we exclude those superfiels
from our consideration. The factor $(1+\x^2)^{-4}$ in the action (B.3) 
was introduced to simplify the transformation properties of the integrand
 under the  $SU(2)_R$ automorphisms of the N=2 supersymmetry algebra, 
as well as the corresponding expressions in N=1 superspace (see below).

It is worth mentioning that the PSS construction above is not
invariant under the $SU(2)_R$ rotations. The $CP(1)$ variable $\x$ has
the rational transformation law,
$$ \x'=\fracmm{\bar{a}\x-\bar{b}}{a+b\x}~~,\eqno(B.7)$$
whose complex $SU(2)_R$ transformation parameters $(a,b)$ are 
constrained by the condition $\abs{a}^2+\abs{b}^2=1$. Nevertheless,
the N=2 supersymmetric action (B.3) is going to be $SU(2)_R$ invariant
too, provided that the function $G$ transforms as 
$$ G(Q',\x')=\fracmm{1}{(a+b\x)^2}G(Q,\x)\eqno(B.8)$$
under the projective transformations (B.7), modulo an additive total 
derivative. Since we have
$$ Q'_{(n)}(\x')=\fracmm{1}{(a+b\x)^n}Q_{(n)}(\x)~,\eqno(B.9)$$
(B.8) puts the severe restriction on a choice of $G$. For example, the 
$SU(2)_R$ invariant PSS Lagrangian of an $O(2)$ multiplet should be 
linear in $Q_2$ outside of the origin of the complex $\x$-plane (in fact, 
up a logarithmical factor), whereas in the case of an $O(4)$ multiplet, its 
$SU(2)_R$ invariant PSS Lagrangian should be proportional to $\sqrt{Q_4}$. 

The invariant action (B.3) of N=2 tensor multiplets $(n=2)$  can be 
easily rewritten to N=1 superspace \cite{klr}. One finds  
$Q_{(2)}=\c-i\x  g +\x^2\bar{\c}$ and
$$ S=\int d^4x d^4\q\fracmm{1}{2\p i}\oint_C d\x\x^{-2}\,
G(\c-i\x g +\x^2\bar{\c},\x) +{\rm h.c.}~,\eqno(B.10)$$
in terms of the N=1 complex {\it chiral\/} superfield 
$\c=\left. L^{11}\right|$ and the N=1 real {\it linear} superfield 
$g=\left. L^{12}\right|$, where $|$ denotes the
$(\bar{\q}_2,\q_1)$-independent part of a superfield or an operator. The N=1
multiplets $\c$ and $g$ together constitute an N=2 tensor multiplet in 4d.
The N=1 superspace covariant derivatives are given by 
$D=\left. D^{2}\right|$ and $\bar{D}=\left.\bar{D}{}^{1}\right|$, whereas the
N=1 superfields $\c$ and $g$ satisfy the constraints
$$ \bar{D}{}^{\dt{\a}}\c=D_{\a}\bar{\c}=0~,\eqno(B.11)$$
and 
$$  \bar{D}_{\dt{\a}}\bar{D}^{\dt{\a}}g=D^{\a}D_{\a}g=0~,\eqno(B.12)$$ 
respectively. The Legendre (duality) transform in N=1 superspace allows 
one \cite{klr} to trade the N=1 linear superfield $g$ for yet 
another N=1 chiral superfield $\j$ ({\it cf.} eq.~(B.26) and the discussion
below). When being applied to the action
(B.10), it yields the dual NLSM action
$$ S=\int d^4x d^4\q\,K(\j+\Bar{\j},\c,\Bar{\c})~,\eqno(B.13)$$
whose hyper-K\"ahler NLSM metric has a K\"ahler potential 
$$ K=\left[ \fracmm{1}{2\p i}  \oint _C d\x \x^{-2}G(\c-i\x H+\x^2\Bar{\c},\x)+
{\rm h.c.}\right] + (\j +\Bar{\j})H~,\eqno(B.14)$$
where the function $H(\c,\Bar{\c},\j+\Bar{\j})$ is a solution to the algebraic
equation
$$\j+\Bar{\j}=\fracmm{1}{2\p}  \oint _C d\x \x^{-1}\fracmm{\pa G}{\pa Q}
(Q,\x)+{\rm h.c.}~,\qquad Q= \c-i\x H+\x^2\Bar{\c}~.\eqno(B.15)$$

As a simple illustration, let's take $G(Q,\x)=F(Q)/\x$, in terms of a {\it
holomorphic} function $F(Q^A)$ of some number $(A)$ of N=2 tensor multiplets, 
with the contour $C$ encircling the origin \cite{ghk99}. Eq.~(B.10) 
yields in this case the N=1 superspace action 
$$\eqalign{
 S[\c,\bar{\c};H]~=~&\int d^4x d^4\q\,\left\{ F_A(\c)\bar{\c}^A -
\frac{1}{2}F_{AB}(\c)H^AH^B +{\rm h.c.} \right\} \cr
~=~& \int d^4x d^4\q\,\left\{ K(\c,\bar{\c})-\frac{1}{2}g\low{AB}(\c,\bar{\c})
H^AH^B\right\},\cr}\eqno(B.16)$$
where we have used the notation
$$ F_A=\fracmm{\pa F}{\pa Q^A}~,\quad F_{AB}=\fracmm{\pa^2 F}{\pa Q^A\pa Q^B}~,
\eqno(B.17)$$
and
$$ K(\c,\bar{\c})=F_A\bar{\c}^A +\bar{F}_A\c^A~,\quad
g\low{AB}(\c,\bar{\c})=F_{AB}(\c)+\bar{F}_{AB}(\bar{\c})~.\eqno(B.18)$$

The N=1 superspace Legendre transform eliminates all the N=1 linear 
superfields $H^A$ in favor of the N=1 chiral
superfields $\j_A$ according to the algebraic equation (B.15),
$$ \j_A + \bar{\j}_A=g\low{AB}(\c,\bar{\c})H^B~, \eqno(B.19)$$
whose solution is 
$$ H^A=g^{AB}(\c,\bar{\c})(\j_A +\bar{\j}_A)~,\eqno(B.20)$$
in terms of the inverse matrix $g^{AB}$. Substituting the solution (B.20)
back into eq.~(B.16) yields a K\"ahler potential in N=1 superspace,
$$K(\c,\bar{\c},\j,\bar{\j})= K(\c,\bar{\c})+
\frac{1}{2}g^{AB}(\c,\bar{\c})(\j_A+\bar{\j}_A)(\j_B+\bar{\j}_B)~.\eqno(B.21)$$
By construction, the K\"ahler potential (B.21) is parametrized by the 
holomorphic potential $F(\c)$, while it yields a hyper-K\"ahler
NLSM metric because of (on-shell) N=2 suspersymmetry. In the context
of the `c-map' (sect.~1), the N=1 chiral superfields $\j$ can be
interpreted as the co-vectors associated with the dual special
K\"ahler manifold \cite{ghk99}.

It is straightforward to reduce the N=2 PSS self-interaction of other 
projective multiplets with $n>2$ in eq.~(B.3) to N=1 superspace. 
For example, the components of the projective $O(4)$ superfield 
$L^{ijkl}$ defined by the constraints (2.29) and (2.30) are given by
$$ L^{ijkl}~;\qquad \l_{\a}{}^{ijk}=D_{\a l}L^{ijkl}~,\quad 
\bar{\l}_{\dt{\a}}{}^{ijk}=\bar{D}_{\dt{\a}l}L^{ijkl}~;$$
$$M^{ij}=-2D_{kl}L^{ijkl}~,\quad \bar{M}^{ij}=-2\bar{D}_{kl}L^{ijkl}~;\qquad
V^{ij}_{\a\dt{\a}}= i\[ \bar{D}_{\dt{\a}k},D\low{\a l}\]L^{ijkl}~;$$
$$\c_{\a k}=D^l_{\a}M_{kl}~,\quad \bar{\c}_{\dt{\a}}{}^k=\bar{D}_{\dt{\a}l}
\bar{M}{}^{kl}~;\qquad C=-2\bar{D}_{ij}D^{kl}L\ud{ij}{kl}~.\eqno(B.22)$$
The fields $(L^{ijkl},\l_{\a}^{ijk})$ are physical,
the fields $(M^{ij},V^{ij}_{\a\dt{\a}})$ are auxiliary, while the fields
$(C,\c^i_{\a})$ play the role of Lagrange multipliers. Varying the
action (B.3) with respect to $C$ yields an algebraic constraint
$${\rm Re}\, \oint_C d\x\,\fracmm{\pa G}{\pa Q }=0 \eqno(B.23)$$
that reduces the number of the independent bosonic degrees of freedom 
on-shell. Indeed, the scalars $L^{ijkl}$ comprise five real bosonic components,
while eq.~(B.23) reduces their number by one, in agreement with the known fact 
that the dimension of a hyper-K\"ahler manifold is always multiple to four.

The N=1 superspace reformulation of the interacting theory (B.3) 
in terms of the projective $O(4)$ multiplet is obtained after a decomposition
of the N=2 extended superfield $L^{ijkl}$ in terms of its N=1 superfield
constituents, 
$$ \left. L^{1111}\right|=\c~,\quad  \left.L^{2222}\right|=\Bar{\c}~,\quad 
4\left.L^{1112}\right|=W~,\quad  6\left.L^{1122}\right|=V~,\eqno(B.24)$$ 
where $\c$ is the N=1 complex chiral superfield satisfying the constraints
(2.1), $W$ is the N=1 {\it complex} linear superfield, 
$\bar{D}_{\dt{\a}}\bar{D}^{\dt{\a}}W=0$, and $V$ is the general N=1 
{\it real} scalar superfield. It gives rise to the N=1 superspace
action \cite{oldket}
$$ S_1=\int d^4x d^4\q\fracmm{1}{2\p i}\oint_C \fracmm{d\x}{\x^2}
G(\c+\x W+\x^2 V-\x^3\Bar{W}+\x^4\Bar{\c},\x)+{\rm h.c.} \eqno(B.25)$$
The N=1 complex linear multiplet $W$ is dual to an N=1 chiral multiplet $\j$, 
while it can be made manifest after introducing the N=1 chiral Lagrange 
multiplier $\j$ into the action (B.25). This yields the master action
$$ S=S_1+\int d^4x d^4\q\,\left(\j\Bar{W}+\Bar{j}W \right)~,\eqno(B.26)$$
where $W$ is now the general complex scalar N=1 superfield. Varying eq.~(B.26)
with respect to $\j$ yields back the constraint on $W$ and the action $S_1$. 
Varying eq.~(B.26) with respect to $W$ instead (i.e. performing the Legendre 
transform) yields an algebraic equation on $W$, which can (at least, 
in principle) be solved in terms of the other superfields. After being 
substituted back into the action (B.26), it results in the dual N=1 
supersymmetric action
$$ S_{\rm dual}=\int d^4x d^4\q\,K(\c,\Bar{\c},\j,\Bar{\j};V)~,\eqno(B.27)$$
with certain function $K$. The general superfield $V$ can now be determined 
from eq.~(B.27), by the use of its algebraic equation of motion. Substituting
the result back into eq.~(B.27) yields the N=2 NLSM hyper-K\"ahler potential 
$K_{\rm h.-K.}(\c,\j;\Bar{\c},\Bar{\j})$ that is only dependent upon the 
N=1 chiral superfields and their conjugates.

As the simplest example, let's consider a free theory defined by 
$$ G(Q_4,\x)\propto \fracmm{Q_4^2}{\x^3}~,\quad {\rm where} \quad
Q_4=\c+\x W+\x^2 V-\x^3\Bar{W}+\x^4\Bar{\c}~,\eqno(B.28)$$
with the contour $C$ encircling the origin in complex $\x$-plane. The 
corresponding N=1 superspace action reads
$$ S_1=\int d^4x d^4\q\,\left(\c\Bar{\c}-W\Bar{W}+V^2\right)~,\eqno(B.29)$$
while its dual chiral (on-shell equivalent) action is given by
$$ S_{\rm dual} =
\int d^4x d^4\q\,\left(\c\Bar{\c}+\j\Bar{\j}\right)~.\eqno(B.30)$$  

A non-trivial example is given by the $O(4)$ tensor multiplet PSS 
selfinteraction parametrized by a holomorphic potential \cite{ghk99}, 
$$ G(Q^A_{(4)};\x) = \fracmm{F(Q^A_{(4)})}{\x^3}~,\eqno(B.31)$$
where $F(Q)$ is a holomorphic function of $Q^A_{(4)}$. Eqs.~(B.3) and
(B.31) lead to the N=1 action
$$\eqalign{
S_1 ~=~ & \int d^4x d^4\q\,\left\{ K(\c,\bar{\c}) 
+g\low{AB}(\c,\bar{\c})\left[ \frac{1}{2}V^AV^B - W^A\bar{W}^B\right]
\right. \cr
 & +\frac{1}{2}\left[ F_{ABC}(\c)W^BW^C +{\rm h.c.}\right]V^A \cr
 & \left.
+ \frac{1}{4!}\left[ F_{ABCD}(\c)W^AW^BW^CW^D +{\rm h.c.}\right]
\right\}~, \cr}\eqno(B.32)$$
where we have used the notation of eqs.~(B.17) and (B.18). The
action (B.32) is quadratic in the general $(V)$ superfields that can
be eliminated according to their algebraic equations of motion,
$$ V^D=-\frac{1}{2}g^{AD}(\c,\bar{\c})\left[ F_{ABC}(\c)W^BW^C + {\rm h.c.}
\right]~. \eqno(B.33)$$
After being substituted back into the action (B.32), it yields the N=1 action
\cite{ghk99}
$$ \eqalign{ S ~=~ & \int d^4x d^4\q \, \left\{ K(\c,\bar{\c}) 
- g_{AB}(\c,\bar{\c}) W^A\bar{W}^B  \right. \cr
 & - \frac{1}{4}g^{EF}(\c,\bar{\c})F_{ABE}(\c)\bar{F}_{CDF}(\bar{\c})
W^AW^B\bar{W}^C\bar{W}^D \cr
& \left.
+\frac{1}{4!}\left[ \cf_{ABCD}(\c,\bar{\c})W^AW^BW^CW^D +{\rm h.c.}\right]
\right\}~, \cr}\eqno(B.34)$$
where we have used the notation
$$ \cf_{ABCD}(\c,\bar{\c})= F_{ABCD}(\c) -3F_{ABE}(\c)g^{EF}(\c,\bar{\c})
F_{CDF}(\c)~.\eqno(B.35)$$
Unfortunately, dualizing the complex linear superfields $W^A$ in favor of
N=1 chiral superfields by the use of the N=1 superfield Legendre transform 
in the general case of eq.~(B.34) does not seem to allow any simple solution.

\section*{Appendix C: N=2 harmonic superspace (HSS)}

In the HSS approach \cite{gikos} one adds harmonics (or twistors) 
$u^{\pm i}$ parametrizing the Riemann sphere $S^2\sim SU(2)/U(1)$ and 
satisfying the relations 
$$ \left( \begin{array}{c} u^{+i} \\ u^{-i}\end{array}\right) \in SU(2)~,\qquad
u^{+i}u^-_i=1~,\quad u^{+i}u^+_i=u^{-i}u^-_i=0~,\quad
\Bar{u^{i+}}=u^-_i~.\eqno(C.1)$$
Then one can make manifest the hidden analyticity structure of the
standard N=2 superspace constraints (subsect.~2.2) defining both N=2 vector
multiplets and Fayet-Sohnius hypermultipets, and find their manifestly 
N=2 supersymmetric solutions in terms of {\it unconstrained} (analytic) 
N=2 superfields, while preserving the $SU(2)_R$ (linearly realized) symmetry.
We follow ref.~\cite{gikos} here.

Instead of using an explicit parametrization of the sphere $S^2$, it
is more convenient to deal with the (equivariant) functions of the
harmonics, which carry a definite $U(1)$ charge defined by 
$U(u^{\pm}_i)=\pm 1$. The simple harmonic integration rules,
$$ \int du =1 \quad{\rm and}\quad \int du\, u^{+(i_1}\cdots
u^{+i_m}u^{-j_1} \cdots
u^{-j_n)}=0 ~~{\rm otherwise}~,\eqno(C.2)$$
are similar to the (Berezin) integration rules in superspace. It is obvious 
that any harmonic integral over a $U(1)$-charged quantity vanishes.

The usual complex conjugation does not preserve analyticity (see
below). However, it does, after being combined with another (star) 
conjugation that only acts on $U(1)$ indices as 
$(u^+_i)^*=u^-_i$ and $(u^-_i)^*=-u^+_i$. One has $ \sbar{u^{\pm i}}
=-u^{\pm}_i$ and
$\sbar{u^{\pm}_i}=u^{\pm i}$. The harmonic covariant derivatives
preserving the defining equations (C.1) are given by
$$ D^{++}=u^{+i}\fracmm{\pa}{\pa u^{-i}}\equiv \pa^{++}~,\quad
D^{--}=u^{-i}\fracmm{\pa}{\pa u^{+i}}~,\quad
D^{0}=u^{+i}\fracmm{\pa}{\pa u^{+i}}-u^{-i}\fracmm{\pa}{\pa u^{-i}}~.
\eqno(C.3)$$
It is easy to check that they satisfy an $su(2)$ algebra and commute
with the standard N=2 superspace covariant derivatives 
$D^i_{\a}$ and $\bar{D}^i_{\dt{\a}}\,$.

The key feature of HSS is the existence of an {\it analytic} subspace 
parametrized by $$ (\z,u)=\left\{ \begin{array}{c}
x^{\m}_{\rm analytic}=x^{\m}-2i\q^{(i}\s^{\m}\bar{\q}^{j)}u^+_iu^-_j~,~~
\q^+_{\a}=\q^i_{\a}u^+_i~,~~ \bar{\q}^+_{\dt{\a}}=\bar{\q}^i_{\dt{\a}}u^+_i~,~~
u^{\pm}_i \end{array} \right\}~,\eqno(C.4)$$
which is invariant under N=2 supersymmetry and is closed under the 
analyticity-preserving conjugation \cite{gikos}. This allows one to 
define {\it analytic} N=2 superfields of any non-negative and integral 
$U(1)$ charge by the analyticity conditions ({\it cf.} the
defining conditions (2.1) of N=1 chiral superfields) 
$$D^+_{\a}\f^{(q)}=\bar{D}^+_{\dt{\a}}\f^{(q)}=0~,\quad {\rm where}\quad
D^{+}\low{\a}=D^i_{\a}u^+_i \quad {\rm and}\quad
\bar{D}^+_{\dt{\a}}=\bar{D}^i_{\dt{\a}}u^+_i~.\eqno(C.5)$$
The analytic measure reads $d\z^{(-4)}du\equiv d^4x^{\m}_{\rm analytic}
d^2\q^+d^2\bar{\q}^+du$ and carries the $U(1)$ charge $(-4)$. The
harmonic derivative $D^{++}$ in the analytic basis (C.4) takes the form
$$D^{++}_{\rm analytic} =\pa^{++}-2i\q^+\s^{\m}\bar{\q}^+\pa_{\m}~,\eqno(C.6)$$
it preserves analyticity, and it allows one to integrate by parts. 
In what follows we omit the explicit reference to the analytic basis,
in order to simplify our notation. 

It is the advantage of HSS that both an off-shell N=2 vector multiplet and an 
off-shell hypermultiplet can be introduced there {\it on equal
footing}. For example, the Fayet-Sohnius hypermultiplet is naturally 
described in HSS by an unconstrained complex analytic superfield $q^+$ 
of $U(1)$ charge $(+1)$. An N=2 vector multiplet in HSS is described
by an unconstrained analytic gauge superfield $V^{++}$ of $U(1)$ charge
$(+2)$. The HSS superfied $V^{++}$ is real with respect to the 
analyticity preserving conjugation, $\Bar{V^{++}}^{\,*}=V^{++}$, while 
it can be naturally introduced as a connection to the harmonic 
derivative $D^{++}$ \cite{gikos}.

An N=2 manifestly supersymmetric free hypermultiplet HSS action 
reads \cite{gikos}~\footnote{Our superfields are all dimensionless, 
whereas the constant $\k$ in front of our actions has dimension 
\newline ${~~~~~}$ of length. We set $\k=1$ for notational simplicity.}
$$S[q]=-\fracmm{1}{\k^2}\int d\z^{(-4)}du\,\sbar{q}{}^+D^{++}q^+
=-\fracmm{1}{2\k^2}\int d\z^{(-4)}du\,q^{a+}D^{++}q^+_a~,\eqno(C.7)$$
where we have introduced the notation $q^+_a=(\sbar{q}{}^+,q^+)$, 
$q^{a+}=\ve^{ab}q^+_b$ and $a=1,2$. Its minimal coupling to an abelian 
N=2 gauge superfield $V^{++}$ reads \cite{gikos}
$$ S[q,V]= - \int d\z^{(-4)}du \,\sbar{q}{}^+(D^{++}+iV^{++})q^+~.
\eqno(C.8)$$
Both actions are also manifestly invariant under the $SU(2)_R$ 
automorphisms, just because of the absence of explicit dependence of
the HSS Lagrangian upon harmonics. The action (C.7), in  fact, possesses the 
extra $SU(2)_{\rm PG}$ internal symmetry rotating the doublet $q^+_a$,
whereas this symmetry is apparently broken in eq.~(C.8) to its $U(1)$ subgroup.
The massless HSS action (C.7) is actually invariant under the full 
N=2 superconformal symmetry isomorphic to $SU(2,2|2)$ that also leaves 
the analytic subspace invariant. The supergroup $SU(2,2|2)$ contains
yet another $SU(2)_{\rm conf.}$ that can be identified on-shell with $SU(2)_R$.

A hypermultiplet (BPS) mass can only come from central charges in N=2 
supersymmetry algebra. The most natural way to introduce central charges 
$(Z,\bar{Z})$ is to identify them with spontaneously broken $U(1)_R$ 
generators of dimensional reduction from six dimensions via the 
Scherk-Schwarz mechanism~\cite{ikz}. After being rewritten to six
dimensions and then `compactified' down to four dimensions, 
the harmonic derivative (C.6) receives an additional `connection' term in 4d,
$$ {\cal D}^{++}=D^{++}+v^{++}~,\quad {\rm where}\quad
v^{++}=i(\theta^+\theta^+)\bar{Z}+i(\bar{\theta}^+\bar{\theta}^+)Z~.
\eqno(C.9)$$
The N=2 central charges can, therefore, be equivalently treated as the
abelian N=2 vector superfield background with the covariantly constant 
N=2 gauge superfield strength $\VEV{W}=Z$. The non-vanishing N=2
central charges break the rigid $U(1)_R$ symmetry,
$ \q^i_{\a}\to e^{-i\g} \q^i_{\a},~\bar{\q}^{\dt{\a}i}\to  e^{+i\g} 
\bar{\q}^{\dt{\a}i}$, of a massless N=2 supersymmetric field theory. 
This symmetry breaking results in the appearance of anomalous terms 
in the N=2 gauge LEEA \cite{sw}, and in the hypermultiplet LEEA as well 
\cite{ikz}.

The general procedure of getting the component NLSM metric from a 
selfinteracting hypermultiplet action in HSS has the following steps 
\cite{gios}:
\begin{itemize}
\item expand the equations of motion in Grassmann variables, and ignore all
the fermionic field components,
\item solve the kinematical differential equations (on the sphere
$S^2\sim SU(2)/U(1)$) for the auxiliary field components, thus eliminating 
the infinite tower of them in the harmonic expansion of the hypermultiplet
HSS superfields,
\item substitute the solution back into the original hypermultiplet 
action in HSS, and integrate over all the anticommuting and harmonic 
coordinates.
\end{itemize}
This most straightforward calculation is very difficult in practice. 
Nevertheless, in was actually done in some special cases of the N=2
NLSM in HSS, including the ones with the four-dimensional Taub-NUT 
\cite{gios}, Eguchi-Hanson \cite{giot} and Gibbons-Hawking (multi-centre)
\cite{ghaw} metrics. 
 
For example, in the Taub-NUT case, the corresponding HSS
hypermultiplet action reads \cite{gios,ikz}
$$ S_{\rm Taub-NUT}[q] =-\int d\zeta^{(-4)}du \left[ \bar{q}^+{\cal D}^{++}q^+ 
+\fracmm{\lambda}{2}(\bar{q}^+)^2(q^+)^2\right]~.\eqno(C.10)$$
The HSS equations of motion for the analytic superfield
$q^+(\zeta,u)$  are given by
$$ {\cal D}^{++}q^+ +\lambda(\bar{q}^+q^+)q^+=0~,\eqno(C.11)$$
where the analytic harmonic derivative ${\cal D}^{++}$ with central charges is
$${\cal D}^{++} = \partial^{++}
-2i\theta^+\sigma^m\bar{\theta}^+\partial_m
+i(\theta^+)^2\bar{Z}+i(\bar{\theta}^+)^2Z~. 
\eqno(C.12)$$
The bosonic terms in the $\theta$-expansion of $q^+$ read \cite{gios}
$$\eqalign{
q^+(\zeta,u)= & F^+(x_{\rm A},u)+i\theta^+\sigma^m\bar{\theta}^+A_m^-
(x_{\rm A},u)+\theta^+\theta^+M^-(x_{\rm A},u)\cr
& +\bar{\theta}^+\bar{\theta}^+ N^-(x_{\rm A},u)+\theta^+\theta^+
\bar{\theta}^+\bar{\theta}^+P^{(-3)}(x_{\rm A},u)~.\cr}\eqno(C.13)$$

The kinematical equations of motion in the $(x^{\m}_{\rm analytic},u)$ 
space, in the presence of central charges, are given by \cite{ikz}
$$\eqalign{
\partial^{++} F^+= & -\lambda(\bar{F}^+F^+)F^+~,\cr
\partial^{++} A^-_m= & 2\partial_mF^+
-\lambda(\bar{F}^+F^+)A^-_m  -\lambda(F^+)^2\bar{A}^-_m~,\cr
\partial^{++} M^{-}= & -\lambda(F^+)^2\bar{N}^-
-2\lambda(\bar{F}^+F^+)M^- -i\bar{Z}F^+~,\cr
\partial^{++} N^- = & -\lambda(F^+)^2\bar{M}^-
-2\lambda(\bar{F}^+F^+)N^- -iZF^+ ~.\cr}
\eqno(C.14)$$

After integrating over the Grassmann variables in the action (C.10) and using
the kinematical equations of motion, one finds that the bosonic action 
reduces to
$$ S_{\rm B}= \frac{1}{2}\int d^4xdu
\left[ A_m^-\partial^m\bar{F}^+ - \bar{A}_m^-\partial^mF^+
- i(\bar{N}^- Z + \bar{M}^- \bar{Z}) F^+ - i\bar{F}^+( Z M^- 
+ \bar{Z} N^-) \right] \eqno(C.15)$$
The kinematical equations for $F^+$ and $A^-_m$ can be easily solved. 
Using the convenient parametrization~\cite{gios}
$$ F^+(x,u)=f^i(x)u^+_i\exp\left[\lambda f^{(j}(x)\bar{f}^{k)}(x)u_j^+u^-_k
\right]~,\eqno(C.16)$$
one finds that
$$ S_{\rm B}= \int d^4x\,\left\{
g_{ij}\partial_mf^i\partial^mf^j+\bar{g}^{ij}\partial_m\bar{f}_i\partial^m
\bar{f}_j + h^i{}_j\partial_mf^j\partial^m\bar{f}_i -V(f)\right\}~,
\eqno(C.17)$$
where the metric is given by~\cite{gios}
$$g_{ij}=\fracmm{\lambda(2+\lambda f\bar{f})}{4(1+\lambda f\bar{f})}
\bar{f}_i
\bar{f}_j~,\quad
\bar{g}^{ij}=\fracmm{\lambda(2+\lambda f\bar{f})}{4(1+\lambda f\bar{f})}
f^if^j~,$$
$$ h^i{}_j=\delta^i{}_j(1+\lambda f\bar{f})
-\fracmm{\lambda(2+\lambda f\bar{f})}{2(1+\lambda f\bar{f})}f^i\bar{f}_j~,
\quad
{\rm and}\quad f\bar{f}\equiv f^i\bar{f}_i~.\eqno(C.18)$$
The metric (C.18) takes the standard Taub-NUT form~\cite{egh}
$$ ds^2=\fracmm{r+M}{2(r-M)}dr^2+\frac{1}{2}(r^2-M^2)
(d\vartheta^2+\sin^2\vartheta d\varphi^2) +2M^2
\left(\fracmm{r-M}{r+M}\right)
(d\psi +\cos\vartheta d\varphi)^2~,\eqno(C.19)$$
after the change of variables~\cite{gios}
$$\eqalign{
f^1=& \sqrt{2M(r-M)}\cos\fracmm{\vartheta}{2}
\exp\fracmm{i}{2}(\psi +\varphi)~,\cr
f^2=& \sqrt{2M(r-M)}\sin\fracmm{\vartheta}{2}
\exp\fracmm{i}{2}(\psi -\varphi)~,\cr}
\eqno(C.20)$$
with
$$f\bar f = 2M(r-M)~, \quad
r\geq M\equiv \fracmm{1}{2\sqrt{\lambda}}~~~,\eqno(C.21)$$
and $M$ being the Taub-NUT mass parameter. 
The non-vanishing auxiliary fields $M^-$ and $N^-$ lead, in addition, 
to a non-trivial scalar potential \cite{ikz},
$$ V(f)=\abs{Z}^2\fracmm{f\bar{f}}{1+ \lambda f\bar{f}}~~.
\eqno(C.22)$$
Further examples can be found in refs.~\cite{giot,ghaw}. The
non-vanishing central charges in the hypermultiplet LEEA may lead to 
spontaneous supersymmetry breaking via the dynamical generation of a
scalar potential, e.g. in the Eguchi-Hanson case \cite{ku}.

\section*{Appendix D: BPS monopoles in the $SU(2)$ 
Yang-Mills-Higgs system, and classical moduli spaces}

The Lagrangian of the $SU(2)$ {\it Yang-Mills-Higgs} (YMH) system in
$1+3$ spacetime
dimensions reads~\footnote{The notation in this Appendix may differ
from the one used in the rest of the paper.}
$$ \Lag_{\rm YMH} =-\fracm{1}{4}\tr(F^2_{\m\n})
+\fracm{1}{2}\tr(D_{\m}\F)^2 -V(\F)~,\eqno(D.1)$$
where both the $SU(2)$ Yang-Mills field $A_{\m}$ and the Higgs field
$\F$ are valued in the Lie algebra of $SU(2)$, 
$$ F_{\m\n}=\pa_{\m}A_{\n}-\pa_{\n}A_{\m}+ie\[A_{\m},A_{\n}\]~,\qquad
D_{\m}\F=\pa_{\m}\F +ie\[A_{\m},\F\]~,\eqno(D.2)$$
with the coupling constant $e$, while the Higgs potential is of the form
$$ V(\F)=\l\,\tr(\F^2-v^2)^2~.\eqno(D.3)$$
A non-vanishing expectaion value, $\VEV{\F}=v$, of the Higgs field breaks 
the gauge symmetry $SU(2)$ down to $U(1)$. The YMH equations of motion
are known to admit the {\it solitonic} solutions given by field 
configurations of finite energy, with the boundary condition $\F^2\to
v^2$ at the special infinity $S^2$ \cite{hp}. The (gauge-inequivalent) 
solitonic solutions form the finite-dimensional space labelled by
integral topological charge $n$ belonging to the homotopy group 
$\p_2[SU(2)/U(1)]=\p_2[S^2]=\p_1[U(1)]={\bf Z}$, which is simply related 
to the magnetic charge $g_{\rm magnetic}=4\p n/e$.

The most {\it fundamental} classical solution, corresponding 
to the static (t'Hooft-Polyakov) monopole \cite{hp} of unit magnetic charge
$(n=1)$, is usually written down in the spherically-symmetric form $(a=1,2,3)$
$$ A^a_i=\ve_{iak}\hat{r}_k\left[\fracmm{1-u(r)}{er}\right]~,\quad
\F^a=\hat{r}_ah(r)~,\eqno(D.4)$$
where $\hat{r}$ is the unit radial vector, while the functions $u(r)$
and $h(r)$ are supposed to satisfy the boundary conditions, $u(0)=1$, 
$u(\infty)=0$, $h(0)=0$ and $h(\infty)=v$, in order to avoid
singularities and have a finite energy. In the so-called 
{\it Bogomol'nyi-Prasad-Sommerfeld} (BPS)
limit \cite{bps}, defined by sending $\l\to 0$ while maintaining  all
the boundary conditions, one finds
$$ u(r)=\fracmm{v}{\sinh(evr)}~,\qquad
h(r)=v\coth(evr)-\fracmm{1}{er}~,\eqno(D.5)$$
which implies for the magnetic field
$$ B^a_i=(D_i\F)^a=\fracmm{\hat{r}_i\hat{r}_a}{er^2} +O(1/r^3)~.\eqno(D.6)$$

A {\it fundamental} (t'Hooft-Polyakov) monopole has four collective 
coordinates, called {\it moduli}. They comprise three translational components,
defining a spacial position of the monopole, and one angular component, 
describing the monopole orientation with respect to the unbroken $U(1)$ 
gauge group and associated with electric charge of the monopole. The
fundamental monopole moduli space is thus given by a non-trivial
bundle $\cm_1=R^3\times S^1$, while it must also be {\it hyper-K\"ahler}
\cite{ati}. The only candidate hyper-K\"ahler metric on $\cm_1$ is
given by the Taub-NUT metric, just because of its isometries and regularity.

The BPS solitons of higher magnetic charge, $n>1$, can be understood
(in asymptotic regions) as being composed of $n$ fundamental monopoles.
In other words, the multi-monopole states should not be considered as
the new states in quantum theory, but they should rather be
interpreted as multi-particle states \cite{wein}. The multi-monopole 
moduli space is also hyper-K\"ahler of real dimension $4n$ \cite{ati}.
The spacial coordinates, describing the `center-of-mass' of a
charge-$n$ monopole configuration, can always be introduced to
factorize the associated $R^3$ factor in the corresponding moduli
space $\cm_n$. The total charge conservation. associated with the
(unbroken) rigid $U(1)$ rotations, further implies the presence of an $S^1$  
factor in $\cm_n$.~\footnote{In the case of higher gauge groups of
rank $r>1$, the total charge may no longer be periodic, \newline ${~~~~~}$
which implies the $R^1$ factor instead of $S^1$ in the decomposition 
(D.7) \cite{lwp}.} The $SU(2)$--based charge-$n$ monopole moduli space 
$\cm_n$ is, therefore, of the form \cite{ati}
$$ \cm_n=R^3\times S^1\times \Tilde{\cm}_n/\G_n~,\eqno(D.7)$$
where $\Tilde{\cm}_n/\G_n$ is called the {\it centered} (or reduced) 
multi-monopole moduli space of real dimension $4(n-1)$, while $\G_n$ stands 
for a discrete subgroup of its isometry group. The $n$-cover $\Tilde{\cm}_n$ of
the centered moduli space is a hyper-K\"ahler manifold too. Its modding by 
$\G_n$ in eq.~(D.7) is necessary to get the right spectrum of quantized 
charges \cite{ati}.

A hyper-K\"ahler metric on the manifold $\Tilde{\cm}_n$ is also supposed to
be regular and complete. In the case of a charge-2 monopole (or, equivalently,
two identical fundamental monopoles), the centered moduli space is of
real dimension four, while it has a rotational isometry $SO(3)$. The 
$\Tilde{\cm}_2$ can thus be identified with the AH manifold.


\begin{thebibliography}{99}

\bibitem{sw} N. Seiberg and E. Witten, \np{426}{94}{19}, \ibid{431}{94}{484}
\bibitem{swg} M. Douglas and S. Shenker, \np{447}{95}{271};\\
A. Klemm, W. Lerche, S. Yankielowicz and S. Theisen, \pl{344}{95}{169};\\
P. Argyres and A. Farragi, \prl{74}{95}{3931};\\
U. Danielson and B. Sundborg, \pl{358}{95}{273};\\
A. Hanany and Y. Oz, \np{452}{95}{283};\\
P. Argyres, M. Plesser and A. Shapere, \np{461}{96}{437};\\
S. Katz, A. Klemm and C. Vafa, \np{497}{97}{173}
\bibitem{rev} A. Bilal, {\it Duality in N=2 SU(2) Yang-Mills theory: a
pedagogical introduction to the work of Seiberg and Witten}, Paris preprint 
LPTENS--95/53; hep-th/9601007;\\
L. Alvarez-Gaum\'e and S. Hassan, Fortschr. Phys. {\bf 45} (1997) 159;\\
W. Lerche, Nucl. Phys. (Proc. Suppl.) {\bf 55B} (1997) 83;\\
P. Di Vecchia, {\it Duality in N=2,4 supersymmetric gauge theories},
NORDITA preprint 98/11--HE; hep-th/9803026;\\
S. V. Ketov, Fortschr. Phys. {\bf 45} (1997) 237
\bibitem{poten} B. de Wit, P. G. Lauwers, R. Philippe, S.-Q. Su and
A. van  Proeyen, \pl{134}{84}{37}
\bibitem{poten2} S. J. Gates, \np{238}{84}{349};\\
S. V. Ketov and I. V. Tyutin, JETP Lett. {\bf 39} (1984) 703 
\bibitem{ind} E. Witten, \np{202}{82}{253}
\bibitem{cfg} S. Cecotti, S. Ferrara and L. Girardello, \ijmp{4}{89}{2475}
\bibitem{polch} J. Polchinski, String Theory, Cambridge University
Press, 1998
\bibitem{sw3} N. Seiberg, \pl{384}{96}{81};\\
N. Seiberg and E. Witten, {\it Gauge dynamics and compactification 
to three dimensions}, in ``Saclay'96. The Mathematical Beauty 
of Physics''; hep-th/9607163 
\bibitem{ch} G. Chalmers and A. Hanany, \np{489}{97}{223}
\bibitem{hwit} A. Hanany and E. Witten, \np{492}{97}{152}
\bibitem{chka} S. A. Cherkis and A. Kapustin, 
Adv. Theor. Math. Phys. {\bf 2} (1998) 1287;
\cmp{203}{99}{713}
\bibitem{adhmn} W. Nahm, \pl{90}{80}{413}; \\
{\it The construction of all self-dual multi-monopoles by the ADHM method}, 
in ``Monopoles in Quantum Field Theory'', eds. N. S. Craigie at al., 
World Scientific, 1982
\bibitem{dancer} A. S. Dancer, Quart. J. Math. Oxford {\bf 45} (1994) 463
\bibitem{chal2} G. Chalmers, Phys. Rev. {\bf D58} (1998) 125011 
\bibitem{dkmtv} N. Dorey, V. Khoze, M. Mattis, D. Tong and
S. Vandoren, \np{502}{98}{59}
\bibitem{lwp} K. Lee, E. Weinberg and P. Yi, \pr{54}{96}{1633}; 6351
\bibitem{gr} G. W. Gibbons and P. Rychenkova, {\it Hyper-K\"ahler
quotient construction of BPS monopole moduli spaces}, DAMTP preprint
R--38--96; hep-th/9608085
\bibitem{ersin} N. Ercolani and A. Sinha, \cmp{125}{89}{385}
\bibitem{hmr} C. J. Houghton, N. S. Manton and N. M. Rom\~ao, {\it On the
constraints defining BPS monopoles}, DAMTP preprint; hep-th/9909168
\bibitem{oldket} S. V. Ketov, Fortschr. Phys. {\bf 36} (1988) 361
\bibitem{klr} A. Karlhede, U. Lindstr\"om and M. Ro\v{c}ek, \pl{147}{84}{297} 
\bibitem{hklr} N. Hitchin, A. Karlhede, U. Lindstr\"om and M. Ro\v{c}ek, 
\cmp{108}{87}{535} 
\bibitem{ok} S. V. Ketov, {\it Self-interaction of N = 2 multiplets in
4d, and the ultra-violet finiteness of 2d, N=4 $\sigma$-models}, in
``Group Theory Methods in Physics'', Nauka, 1985, p.~87
\bibitem{gikos} A. S. Galperin, E. A. Ivanov, V. I. Ogievetsky, 
S. Kalitzin and E. Sokatchev, \cqg{1}{84}{469}
\bibitem{fre} P. Fr\'e and P. Soriani, {\it N=2 Wonderland}, World Scientific,
1995, Ch.~8
\bibitem{hk1} S. V. Ketov, \pl{399}{97}{83}
\bibitem{ikz} E. A. Ivanov, S. V. Ketov and B. M. Zupnik, \np{509}{98}{53}
\bibitem{rev2}  S. V. Ketov, Fortschr. Phys. {\bf 47} (1999) 643
\bibitem{manton} N. S. Manton, \pl{110}{82}{54}
\bibitem{ku} S. V. Ketov and Ch. Unkmeir, \pl{422}{98}{179}
\bibitem{sspace} S. J. Gates, Jr., M. T. Grisaru, M. Ro\v{c}ek and W.
 Siegel, {\it Superspace or One Thousand and One Lessons in Supersymmetry}, 
Benjamin/Cummings, 1983;\\ 
I. L. Buchbinder and S. M. Kuzenko, {\it Ideas and Methods 
of Supersymmetry and Supergravity}, IOP Publishers, 1998
\bibitem{haklr} C. M. Hull, A. Karlhede, U. Lindstr\"om and M.
 Ro\v{c}ek, \np{266}{86}{1} 
\bibitem{gios} A. S. Galperin, E. A. Ivanov, V. I. Ogievetsky 
and E. Sokatchev, \cmp{103}{86}{515}
\bibitem{gios2} A. S. Galperin, E. A. Ivanov, V. I. Ogievetsky 
and E. Sokatchev, Ann. Phys. (N.~Y.) {\bf 185} (1988) 1; 22  
\bibitem{zumino} B. Zumino, \pl{87}{79}{203}
\bibitem{difg} S. Kobayashi, K. Nomizu, {\it Foundations of Differential 
Geometry}, Interscience, 1969 
\bibitem{sig} E. Sokatchev, \np{99}{75}{96};\\
S. J. Gates, Jr. and W. Siegel, \np{189}{81}{295}
\bibitem{wess} J. Wess, Acta Phys. Austr. {\bf 41} (1975) 409
\bibitem{fas}  P. Fayet, \np{113}{76}{135};\\
M. F. Sohnius, \np{138}{78}{109}
\bibitem{ati} M. Atiyah and N. Hitchin, {\it The Geometry 
and Dynamics of Magnetic Monopoles}, Princeton University Press, 1988;
Phys. Lett. {\bf 107A} (1985) 21; Phil. Trans. R. Soc. Lon. {\bf A315}
(1985) 459
\bibitem{flag} G. G. Hartwell and P. S. Howe, Int. J. Mod. Phys. {\bf A10}
(1995) 3901; \cqg{12}{95}{1823};\\
P. Howe, {\it On harmonic superspace}, contributed to 
International Seminar on Supersymmetries and Quantum Symmetries in Dubna, 
Russia, 22--26 July, 1997; London preprint KCL--TH--98--62; hep-th/9812133 
\bibitem{kuzp} S.M.~Kuzenko,  Int. J. Mod. Phys. {\bf A14} (1999) 1737
\bibitem{lroc} U. Lindstr\"om and M. Ro\v{c}ek, \cmp{115}{88}{21}; 
ibid. {\bf 128} (1990) 191   
\bibitem{bag} J. Bagger, \np{211}{83}{302}
\bibitem{beg} J. Bagger, A. S. Galperin, E. A. Ivanov, V. I. Ogievetsky,
\np{303}{88}{522}
\bibitem{hssf} A. S. Galperin, E. A. Ivanov, V. I. Ogievetsky and E. Sokatchev,
\cqg{2}{85}{601}; 617
\bibitem{gio} A. S. Galperin, E. A. Ivanov and V. I. Ogievetsky, 
\np{282}{83}{74}
\bibitem{prep} S. V. Ketov, \pl{469}{99}{136}
\bibitem{iro} I. T. Ivanov amd M. Ro\v{c}ek, \cmp{182}{96}{291}
\bibitem{don} S. Donaldson, \cmp{96}{84}{387}
\bibitem{oli} D. Olivier, Gen. Rel. and Grav. {\bf 23} (1991) 1349
\bibitem{hur} J. Hurtubise, \cmp{92}{83}{195}, \ibid{100}{85}{191}
\bibitem{bakas} I. Bakas, {\it Remarks on the Atiyah-Hitchin metric}, 
Patras preprint; hep-th/9903256
\bibitem{fkra} H. Farkas and I. Kra,  {\it Riemann Surfaces}, 
Springer--Verlag, 1980
\bibitem{klnt} M. Ko, M. Ludvigsen, E. T. Newman and K. Tod,
Phys. Rep. {\bf 71} (1981) 51
\bibitem{gm} G. Gibbons and P. Ruback, \cmp{115}{88}{267}
\bibitem{wein} E. J. Weinberg, {\it Massive and massless monopoles and
duality}, Columbia preprint CU--TP--946; hep-th/9908095
\bibitem{gno} P. Goddard, J. Nuyts and D. Olive, \np{125}{77}{1} 
\bibitem{chal} G. Chalmers, {\it Multi-monopole moduli spaces for SU(N)
gauge group}, Stony Brook preprint, ITP--SB--96--12; hep-th/9605182
\bibitem{hitchin} N. J. Hitchin, \cmp{83}{82}{579}; ibid. {\bf 89} (1983) 145
\bibitem{book} S. V. Ketov, {\it Conformal Field Theory}, World Scientific, 
1995
\bibitem{hou} C. J. Houghton, {\it On the generalized Legendre transform and
monopole metrics}, Columbia preprint; hep-th/9910212
\bibitem{jarvis} S. Jarvis, {\it A rational map for euclidean
monopoles via radial scattering}, Oxford preprint, 1996
\bibitem{mhpa} Th. Ioannidou and P. M. Sutcliffe, {\it Monopoles and
harmonic maps}, Kent preprint UKC/IMS/99/07; hep-th/9903183
\bibitem{dwi} R. Donagi and E. Witten, \np{460}{96}{299}
\bibitem{w} E. Witten, \np{500}{97}{3}
\bibitem{besse} A. Besse, {\it Einstein Manifolds}, Springer--Verlag, 1987
\bibitem{hst} P. S. Howe, K. S. Stelle and P. K. Townsend, \np{214}{83}{519}
\bibitem{ghk99} S. J. Gates Jr., T. H\"ubsch and S. M. Kuzenko, 
\np{557}{99}{443}
\bibitem{giot} A. S. Galperin, E. A. Ivanov, V. I. Ogievetsky and
P. K. Townsend. Class. Quantum Grav. 3 (1986) 625
\bibitem{ghaw} G. W. Gibbons, D. Olivier, P. J. Ruback and G. 
 Valent, \np{296}{88}{679}
\bibitem{egh} T. Eguchi, P. Gilkey and A. Hanson, Phys. Rep. 
{\bf 66} (1980) 213
\bibitem{hp} G. t'Hooft, \np{79}{74}{276};\\
A. M. Polyakov, JETP Lett. {\bf 20} (1974) 194
\bibitem{bps} E. B. Bogomol'nyi, Sov. J. Nucl. Phys. {\bf 24} (1976) 449;\\
M. K. Prasad and C. H. Sommerfeld, Phys. Rev. Lett. {\bf 35} (1975) 760.

\end{thebibliography}
\end{document}